\documentclass[showpacs,prb,twocolumn,floats,superscriptaddress]{revtex4}
\usepackage{graphicx}
\usepackage{amsmath}
\usepackage{amssymb}
\usepackage[latin1]{inputenc}
\usepackage[french]{babel}
\usepackage{latexsym}

\usepackage{graphicx} 
\usepackage{float}
\usepackage{tabularx}
\usepackage{color}
\usepackage{array}
\usepackage{float}
\usepackage[colorlinks]{hyperref}

\begin{document}

\title{Band structure and Klein paradox for a pn junction in ABCA-tetralayer graphene}
\date{\today }
\author{Abderrahim EL MOUHAFID}
\email{elmouhafid@gmail.com}
\affiliation{Laboratory of Theoretical Physics, Faculty of Sciences, Choua\"ib Doukkali University, PO Box 20, 24000 El Jadida, Morocco}
\author{Ahmed JELLAL}
\email{a.jellal@ucd.ac.ma}
\affiliation{Laboratory of Theoretical Physics, Faculty of Sciences, Choua\"ib Doukkali University, PO Box 20, 24000 El Jadida, Morocco}
\affiliation{Saudi Center for Theoretical Physics, Dhahran, Saudi
Arabia}

\begin{abstract}

We investigate the band structure of ABCA-tetralayer graphene (ABCA-TTLG) subjected 
to an external potential 
$V$ applied between top and bottom layers. Using
the tight-binding model, including the nearest $t$ and next-nearest-neighbor $t'$ hopping, 
low-energy model and two-band approximation model we study 
the band structure variation
along the  lines $\Gamma-M-K-\Gamma$ in the first Brillouin zone, electronic band gap near
Dirac point
 $K$   and transmission properties, respectively. Our results reveal that  ABCA-TTLG 
exhibits markedly different properties as functions of $t'$ and $V$. 
We show that the hopping parameter $t'$ changes the energy dispersion,   
the position of  $K$  and breaks sublattice symmetries. 
A sizable band gap is created at $K$,  
which could be opened and controlled by the applied potential $V$.  
This gives rise to 1D-like van Hove 
singularities (VHS) in the  density of states (DOS). We study  
the relevance of the skew hopping parameters $\gamma_3$ and $\gamma_4$ to these properties and show that for 
energies $E\gtrsim6$meV their effects are negligible. Our results are numerically discussed 
and compared with the  literature.
\end{abstract}

\maketitle
\section{Introduction}
 Generally, graphene can be stacked in various ways to form  multilayered graphenes 
 (MLG) with  different physical properties. Typical graphene stacking includes Order, Bernal (AB), 
 and Rhombus stacking (ABC)\cite{Aoki2007,Dresselhaus200286,Jhang11, Koshino2010, Lui11,Koshino200923}. 
 In Order stacking, all carbon atoms of each layer are well-aligned. For AB and ABC stacking,
 a cycle period is constituted by two layers and three layers of non-aligned graphene, respectively. 
 It has been showed  that the properties of graphene like band structure, band gap, transport properties, 
 optical properties and density of state depend on the way how graphene is stacked 
 \cite{Latil2006,Mikito2010,Aoki2007,Cocemasov2013,Mak2010,Guinea200626,Avetisyan200901,
 Avetisyan201032,Lu200627,Ben201301,Chegel201683,Sheng200966}
 and also on  the application of external sources 
 \cite{Aoki2007,Mikito2010,Kumar201101}. Indeed, Aoki and Amawashi 
 showed that the AB-stacked MLG is a semi-metal with an electrically tunable 
 band overlap, while the ABC-stacked MLG is a semiconductor with an electrically 
 tunable band gap\cite{Aoki2007}. 
 Mak {\it et al.} experimentally investigated
 the electronic structure of few-layer graphene samples with crystalline order 
  by infrared absorption spectroscopy for photon energies \cite{Mak2010}.
 Lu {\it et al.}
 investigated the influence of AB stacking on the optical properties of MLG in an electric 
 field\cite{Lu200627}. Ben {\it et al.} studied the influence of ABC stacking on 
 the Klein and anti-Klein tunneling  of MLG in an external potential \cite{Ben201301}. 
 As a consequence, MLG 
 exhibits the rare behavior of crystal structure modification, 
 and hence modification of electronic properties, via the application of the potential, 
  electric and magnetic fields.

In the present work, we study  the electronic properties of 
 ABCA-TTLG in the presence of an external 
 potential 
$V$ applied between top and bottom layers. These properties are explored by employing 
the tight-binding Hamiltonian,  low-energy and  two-band 
approximation models. They are strongly dependent on the geometric structure 
and the applied potential  $V$. The application of $V$ remarkably modifies 
the energy dispersions, causes the subbands anticrossing, changes the subbands
spacing and induces the oscillating bands. Here 
we are limited ourselves to next-nearest-neighbor hoppings $t'$ and neglected other 
less important hoppings that are present in  ABCA-TTLG. 
We show that  
$t'$  affects  the density of states (DOS) 
of ABCA--TTLG. 
At low energy the Klein (KT) and anti-Klein (AKT) tunneling
are analyzed.

The manuscript is organized as follows. In Sec. \ref{Hmodel}, we  introduce   
the tight-binding Hamiltonian,  low-energy  and two-band approximation 
models to calculate the band structure of  ABCA--TTLG. 
Sec. \ref{Bandstructure} and \ref{trans} are devoted 
to numerical analysis  and discussion of 
the band structure behavior and 
  transmission probabilities for electrons impinging on a potential step (pn junction). 
Our main conclusions are summarized in Sec. \ref{Conclusion}.
\section{The Hamiltonian model}
\label{Hmodel}
\begin{figure}[tb]
\centering
\includegraphics[width=6cm]{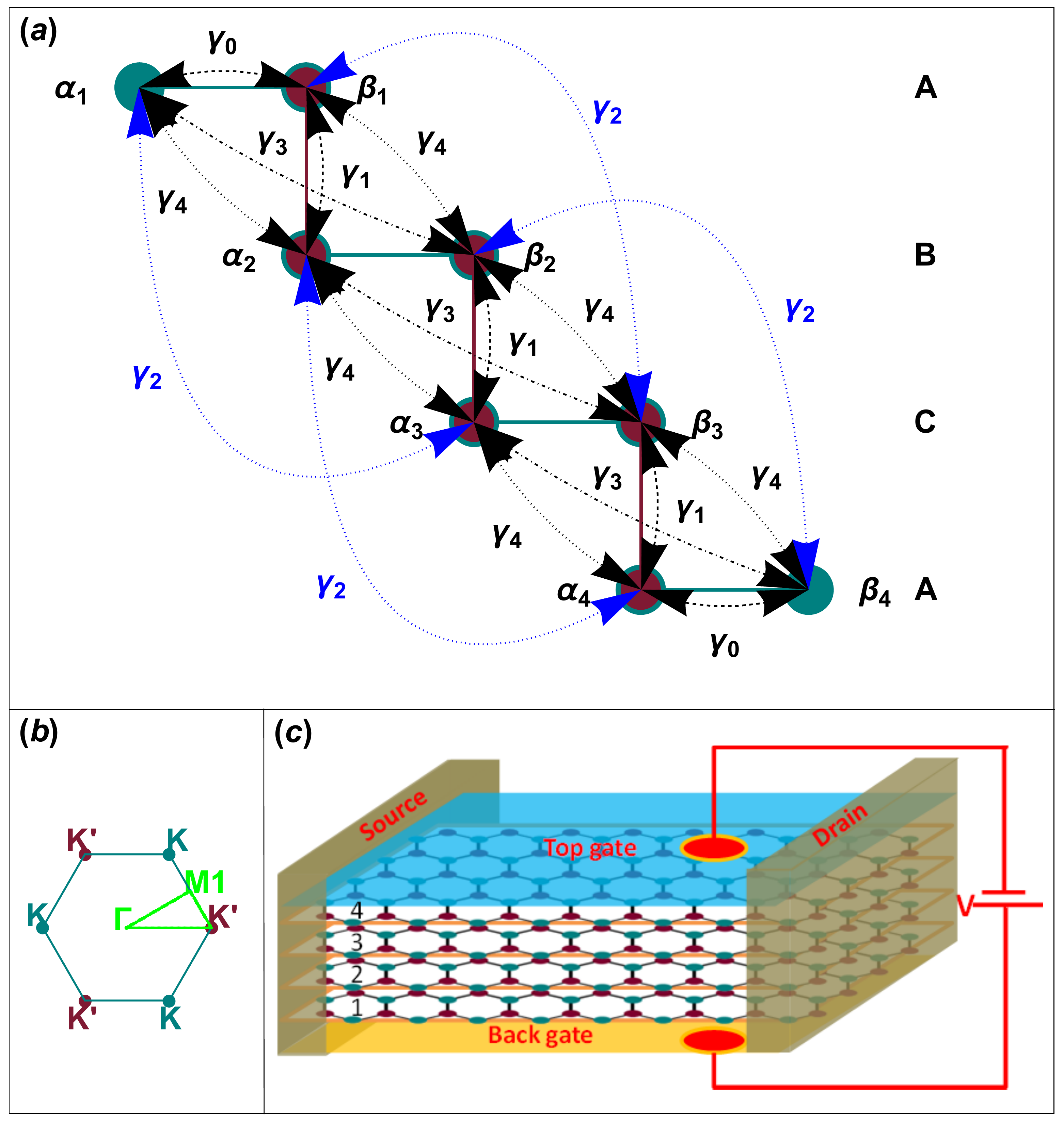}
\caption{(Color online) (a) Tight-binding diagram for four stacking 
types ABCA-TTLG showing the Slonczewski-Weiss-McClure parameterization \cite{Dresselhaus200286} 
of relevant couplings $\gamma_{0}$ to $\gamma_{4}$. (b) Schematic of the hexagonal Brillouin zone (BZ) 
with two inequivalent valleys $K$ and $K'$. (c) Schematic representation of ABCA-TTLG junction. 
There is an electrostatic potential difference  $V=2U_{0}$ between top and bottom layers. 
Sublattices $\alpha_i$ and $\beta_i$ are represented in different colors.}\label{stackingTTL}
\end{figure}
We consider
a lattice of ABCA-TTLG consists of four coupled layers\cite{Koshino2010,Ben201301}, each with carbon atoms arranged 
on a honeycomb lattice, including pairs of inequivalent sites 
$\{\alpha_{1},\beta_{1}\}$, $\{\alpha_{2},\beta_{2}\}$, $\{\alpha_{3},\beta_{3}\},$ 
and $\{\alpha_{4},\beta_{4}\}$ in the top, center, and bottom layers, respectively. 
The layers are arranged as shown in Fig. \ref{stackingTTL}(a) such that pairs of sites 
$\beta_{1}$ and $\alpha_{2}$, $\beta_{2}$, $\alpha_{3}$ and $\beta_{3}$ and $\alpha_{4}$, 
lie directly above or below each others. 
 The different sublattices $\alpha_{i}$ and $\beta_{i}$ are represented by darkred 
 and teal  solid balls, respectively. In order to write down an effective mass Hamiltonian, 
 we adapt the Slonczewski-Weiss-McClure parameterization of tight-binding couplings 
 of bulk graphite \cite{Dresselhaus200286}. We include parameters 
 $\gamma_{0}$, $\gamma_{1}$, $\gamma_{2}$, $\gamma_{3}$ and $\gamma_{4}$, 
 where $\gamma_{0}$ 
 the intralayer coupling 
 $\alpha_{i}\leftrightarrow\beta_{i}$ ($i=1,2,3,4$), $\gamma_{1}$ 
 the interlayer coupling $\beta_{i}\leftrightarrow\alpha_{i+1}$ ($i=1,2,3$), $\gamma_{2}$ 
 the interlayer coupling $\alpha_{i}\leftrightarrow\alpha_{i+2}$ 
 and $\beta_{i}\leftrightarrow\beta_{i+2}$ ($i=1,2$), $\gamma_{3}$ 
 the interlayer coupling between  $\alpha_{i}$ and  $\beta_{i+1}$ ($i=1,2,3$), 
  $\gamma_{4}$ the interlayer coupling $\alpha_{i}\leftrightarrow\alpha_{i+1}$ 
 and $\beta_{i}\leftrightarrow\beta_{i+1}$ ($i=1,2,3$). For typical values, we quote \cite{Dresselhaus200286} 
 $\gamma_{0}\approx3.16$eV,$\gamma_{1}\approx0.39$eV, $\gamma_{2}\approx-0.020$eV, $\gamma_{3}\approx0.315$eV 
 and $\gamma_{4}\approx0.044$eV. These interatomic coupling parameters are depicted in Fig. \ref{stackingTTL}(a). 
 The contribution of the skew hopping parameter $\gamma_3$ results in 
 the so called\cite{McCann2006, McCann2007} trigonal warping, an effect occurring only at very 
 low energy ($E\lesssim 6meV$) which will be discussed  
 in Sec 
 \ref{Lowenergy}. The parameter $\gamma_4$ has an even lower impact on the electronic properties,
 see Sec \ref{Lowenergy}. Therefore, we will often neglect these two 
 $\gamma$-parameters in Sec \ref{TBModel} and \ref{Twoband}. There is a degeneracy point 
 at each of two inequivalent corners $K$ and $K'$ of the hexagonal first Brillouin zone, also referred 
 to as valleys, Fig. \ref{stackingTTL}(b). Near the centre of each valley, there are eight electronic bands. 
 In the basis  $\{\alpha_{1}, \beta_{1}, \alpha_{2}, \beta_{2}, \alpha_{3},  \beta_{3}, \alpha_{4}, \beta_{4}\}$ 
 the electronic properties for ABCA-TTLG are then obtained from the following Hamiltonian matrix\cite{Koshino2010}
\begin{eqnarray}
\mathcal{H}(\mathbf{k})=\left[
\begin{array}{ccccc}
\mathcal{H}_{1}(\mathbf{k}) & \Gamma_{1} & \Gamma_{2} & 0 \\
\Gamma_{1}^{\dag } & \mathcal{H}_{2}(\mathbf{k}) & \Gamma_{1} & \Gamma_{2} \\
\Gamma_{2}^{\dag } & \Gamma_{1}^{\dag } & \mathcal{H}_{3}(\mathbf{k}) & \Gamma_{1} \\
0 & \Gamma_{2}^{\dag } & \Gamma_{1}^{\dag } & \mathcal{H}_{4}(\mathbf{k})%
\end{array}%
\right]\label{HamABCA},
\end{eqnarray}%
where the interlayer couplings $\Gamma_{1}$ and $\Gamma_{2}$ are given by
\begin{eqnarray}
\Gamma_{1}=\left[
\begin{array}{cc}
-v_{4}f(\mathbf{k}) & v_{3}f^{\dagger}(\mathbf{k}) \\
\gamma_{1} & -v_{4}f(\mathbf{k})%
\end{array}%
\right]\label{Gamma2}, & \mbox{} \Gamma_{2}=\left[
\begin{array}{cc}
0 & \gamma_{2}/2 \\
0 & 0%
\end{array}%
\right]\label{Gamma2}
\end{eqnarray}%
with $v_{3,4}=v_{F}\gamma_{3,4}/\gamma_{0}$ are related to the skew hopping parameters, 
$v_{F}=3a\gamma _{0}/(2\hslash)$ is the Fermi velocity in terms of the in-plane  nearest neighbor hopping 
$\gamma_{0}$,  $a$ is the lattice constant. The  $2\times2$ matrix $\mathcal{H}_{i}(\mathbf{k})$ 
describes the intralayer processes for ABCA-TTLG is
\begin{eqnarray}
\mathcal{H}_{i}(\mathbf{k})=\begin{pmatrix} V_{i} & f(\mathbf{k}) \\
f^{\dagger}(\mathbf{k}) & V_{i} \\
\end{pmatrix}, & \mbox{} 
\end{eqnarray}%
$i=1, 2, 3, 4$ labeling the graphene layers,
 $V_{i}$ describe an external potential in each layers, $f(\mathbf{k})$ 
is a momentum space representation of intersublattice hopping processes for electrons  
with wave vector $\mathbf{k}$. In the framework of a  tight-binding approximation 
(see Sec. \ref{TBModel}),  $f(\mathbf{k})$ is given by\cite{Hasegawa200613,Dietl200805,Wunsch200827}
\begin{eqnarray}
f(k)=-t'-t\left(e^{-ik.a_1}+e^{-ik.a_2}\right)\label{HTB},
\end{eqnarray}%
where $a_1=(a/2)(-\sqrt{3}e_{x}+e_y)$ and $a_2=(a/2)(\sqrt{3}e_{x}+e_y)$ are 
basis vectors of the triangular Bravais lattice, in terms of the lattice constant $a$, $t$ is 
the nearest-neighbor hopping energy (hopping between different sublattices), and $t'$ 
the next nearest-neighbor hopping integral (hopping in the same sublattice). The angular dependence 
of $f(k)$ in Eq. \ref{HTB} is called trigonal warping because it leads to a deformation of the form of 
the Fermi line around the centre of each valley. This deformation increases with an increase of 
the absolute value of the wave vector. In ABCA-TTLG, a second cause of trigonal warping is
the parameter $\gamma_3$ describing direct interlayer coupling between  $\alpha_{i}$ and  
$\beta_{i+1}$($i=1,2,3$), leading to an effective velocity $v_{3}=v_{F}\gamma_{3}/\gamma_{0}$. 
In the low-energy  model (see Sec. \ref{Lowenergy}),  $f(\mathbf{k})$ is 
\begin{eqnarray}
f(\mathbf{k})=\hbar v_{F}(k_{x}-ik_{y}),\label{hamDirac}
\end{eqnarray}%
where $\mathbf{k}=(k_{x},k_{y})$ is the two dimensional momentum operator. 
The applied potential $V_{i}$  
can be varied by gating the sample with top and back gates (see Fig. \ref{stackingTTL}(c)). This is
\begin{eqnarray}
V_{i}=\begin{cases}U_{0} & \mbox{for } i=1 \mbox{ (layer 1)} \\
0 & \mbox{for } i=2  \mbox{ (layer 2)} \\
0 & \mbox{for } i=3  \mbox{ (layer 3)} \\
-U_{0} & \mbox{for } i=4  \mbox{ (layer 4)}, \\
\end{cases}\label{potential}
\end{eqnarray} where $U_{0}$ describes a 
potential difference between layers 1 and 4.
\begin{figure*}[tbh]
\begin{center}
\end{center}
\includegraphics[width=1.7in]{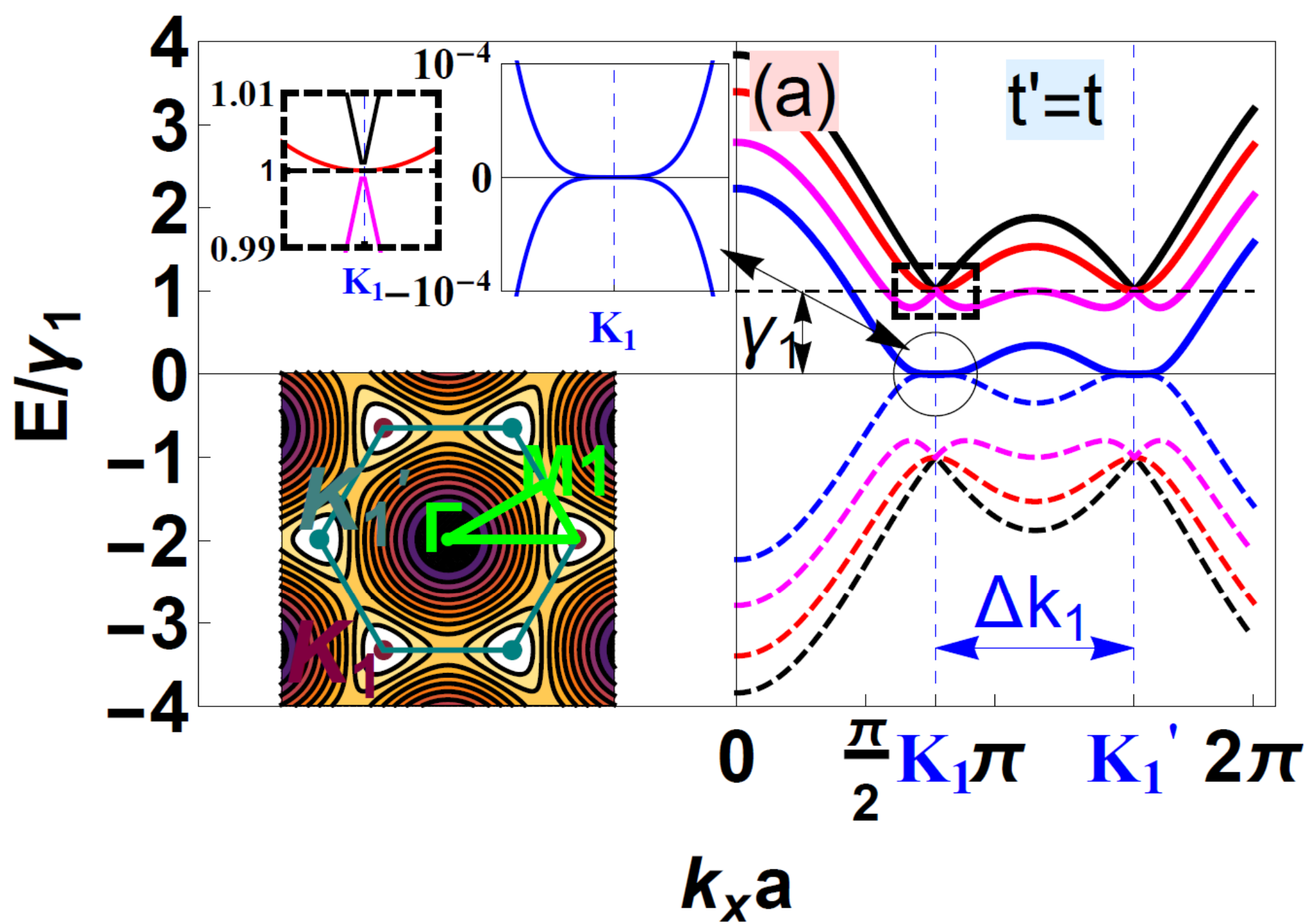}\includegraphics[width=1.7in]{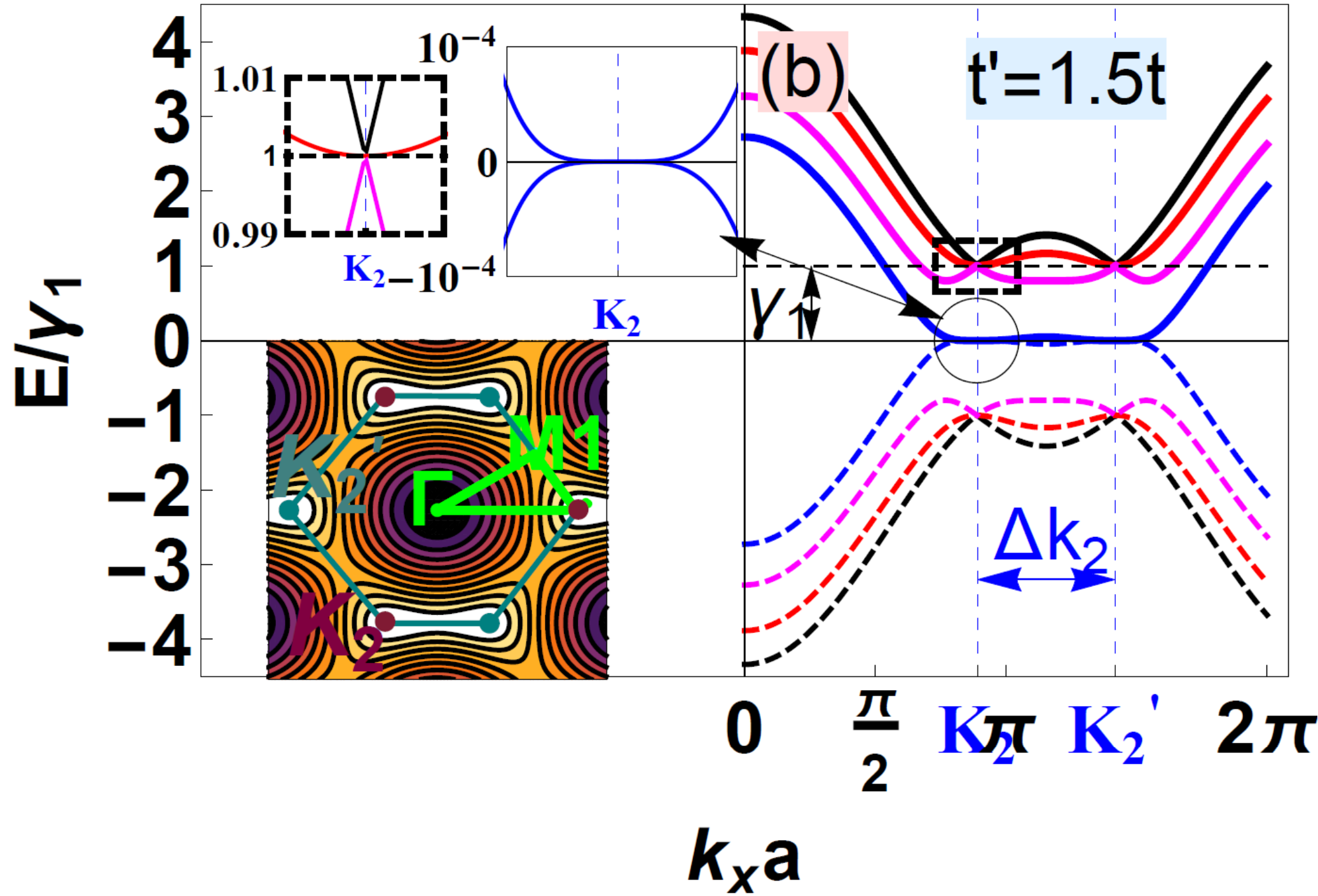}
\includegraphics[width=1.7in]{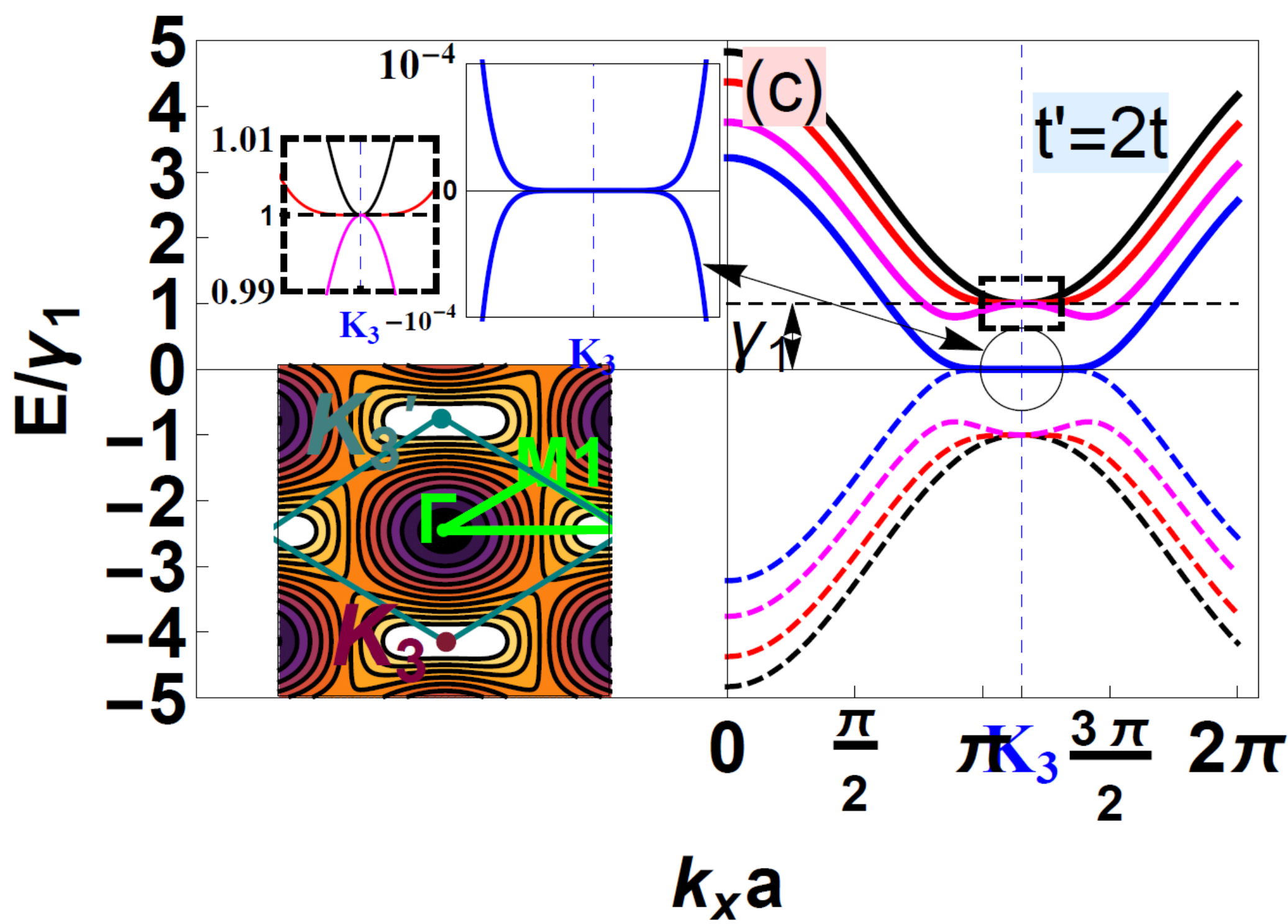}\includegraphics[width=1.7in]{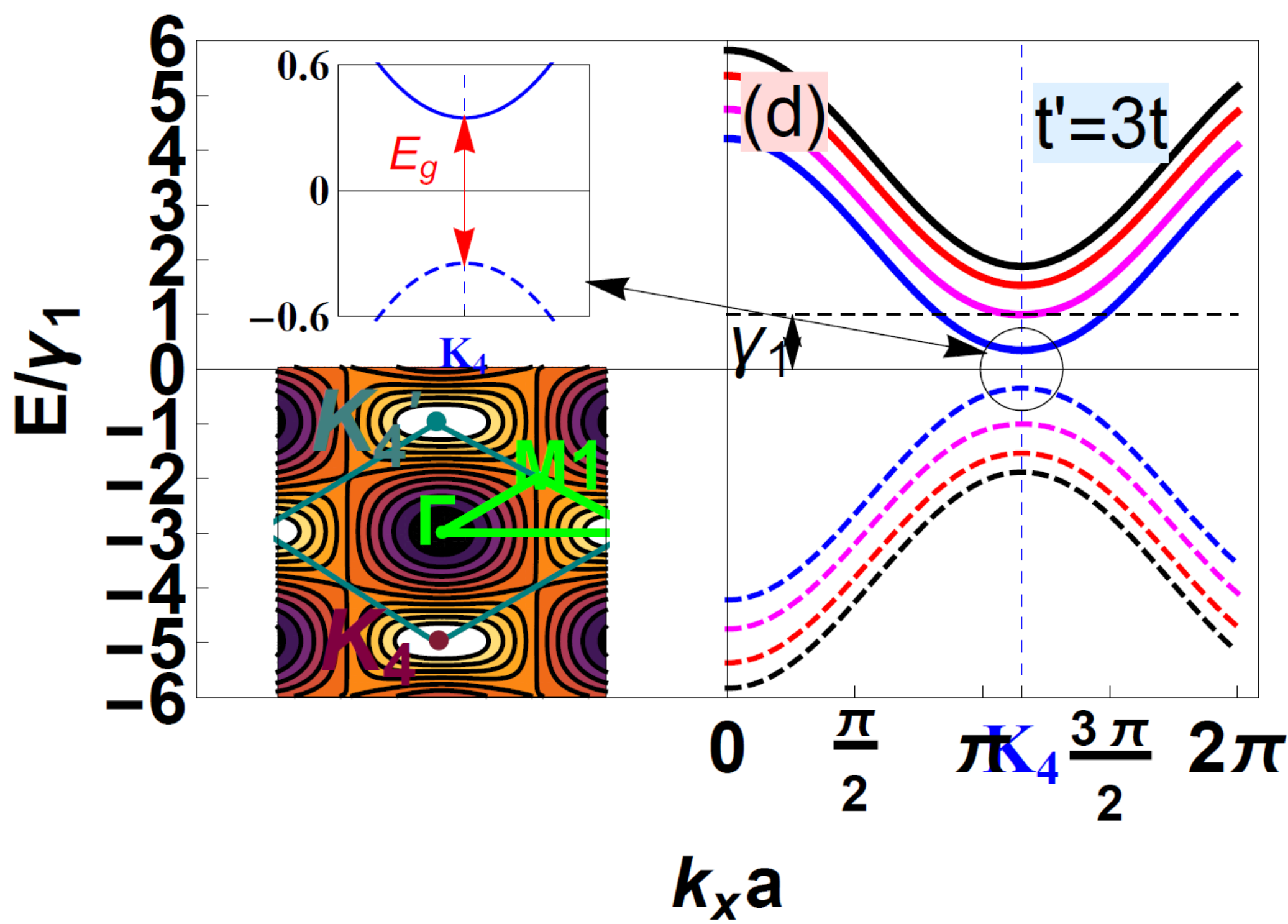}
\caption{(Color online) Band structure of ABCA-TTLG for $k_y=0$ and $U_{0}=0$. 
The blue, magenta, red and black curves correspond to the tight-binding model. 
The solid curves correspond to the conduction bands and the dashed curves correspond
to the valance bands. (a) $t'=t$, (b) $t'=1.5t$, (c) $t'=2t$ and (d) $t'=3t$. 
The vertical dashed lines indicates the position of the  points $K$.}
\label{energyeffec1}
\end{figure*}
\begin{figure}[tb]
\centering
\includegraphics[width=6.9cm]{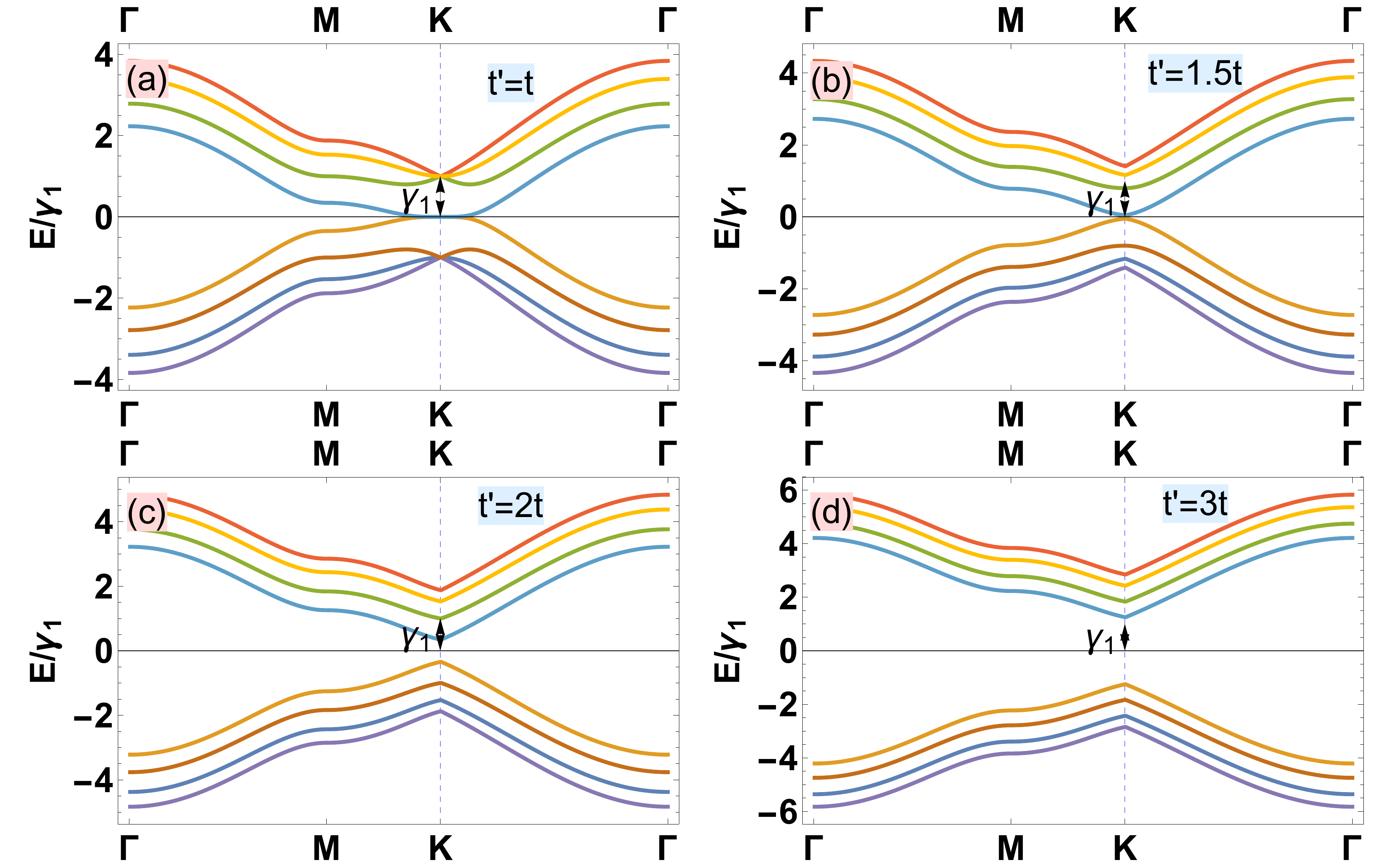}
\caption{(Color online) Band structure of ABCA-TTLG along the lines $\Gamma-M-K-\Gamma$ 
 in the first BZ for $U_{0}=0$, (a) $t'=t$, (b) $t'=1.5t$, (c) $t'=2t$, (d) $t'=3t$. $K$, 
$\Gamma$ and $M$ are, respectively, the corner, the center and the middle point between two corners 
in the hexagonal BZ as shown in Fig. \ref{stackingTTL}(b).}\label{EnergyGMKG}
\end{figure}

\section{Band structure}
\label{Bandstructure}
\subsection{Tight-binding model}
\label{TBModel}
\begin{figure}[tb]
\centering
\includegraphics[width=2.9cm]{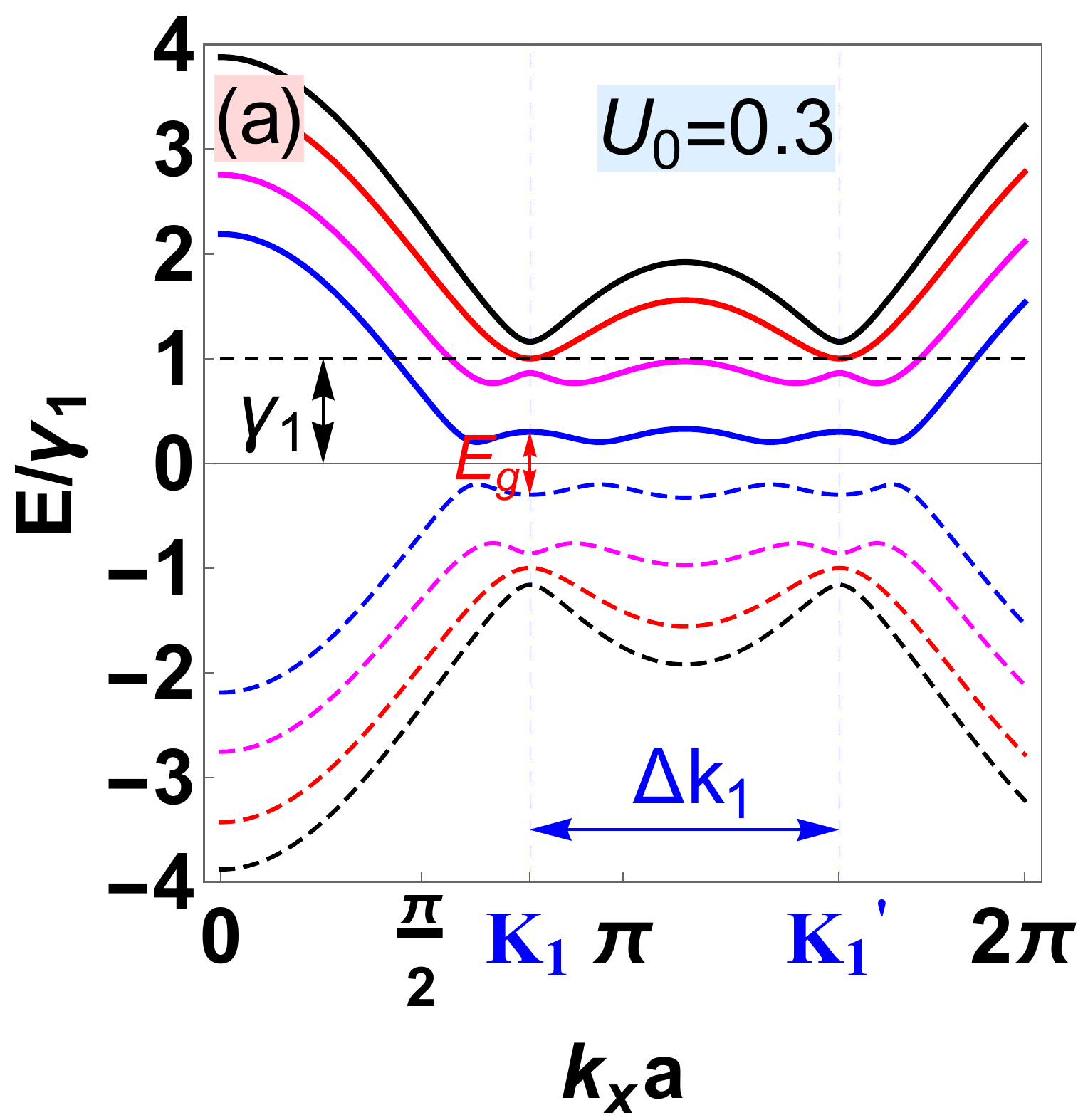}\includegraphics[width=2.9cm]{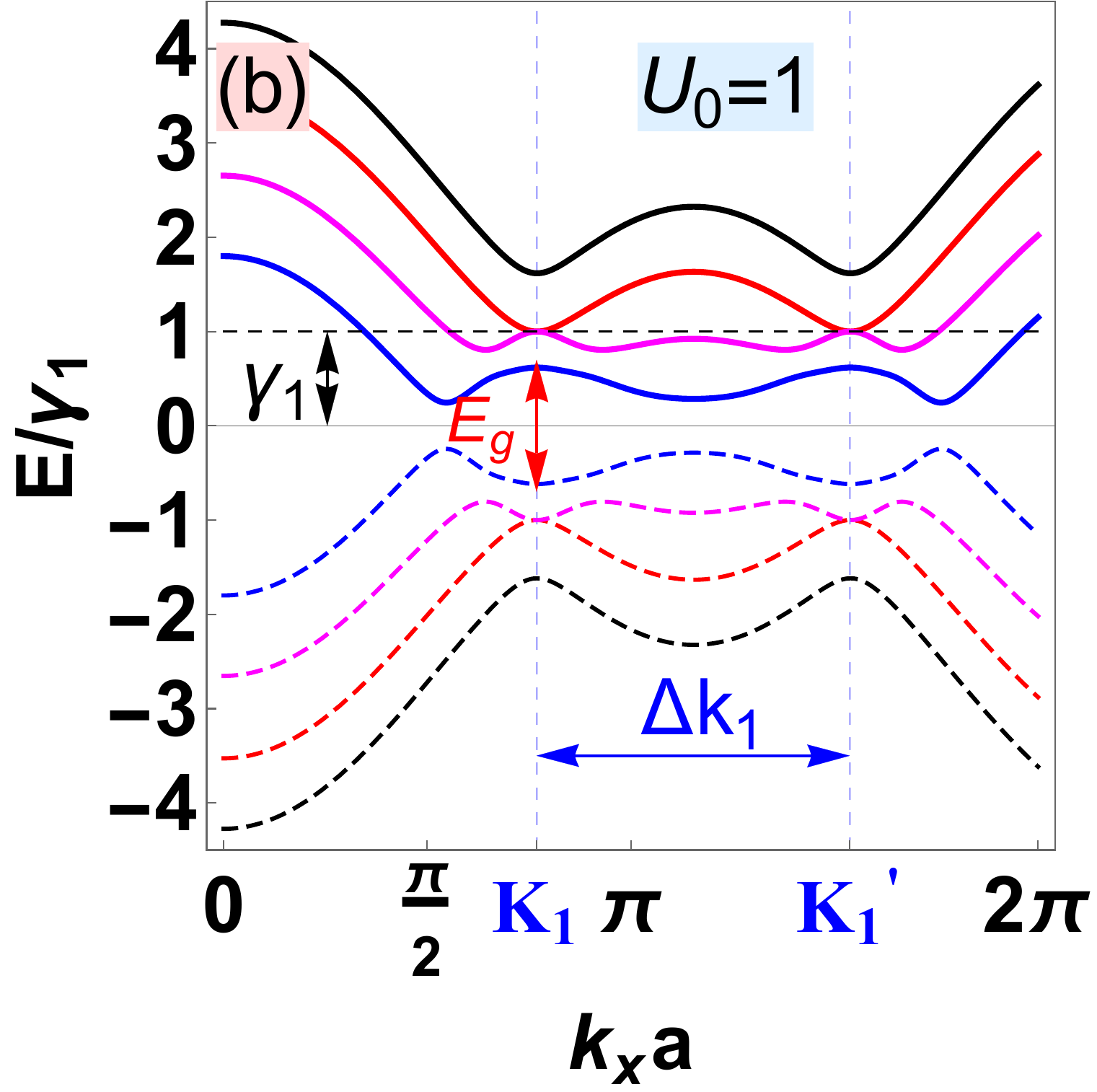}\includegraphics[width=2.9cm]{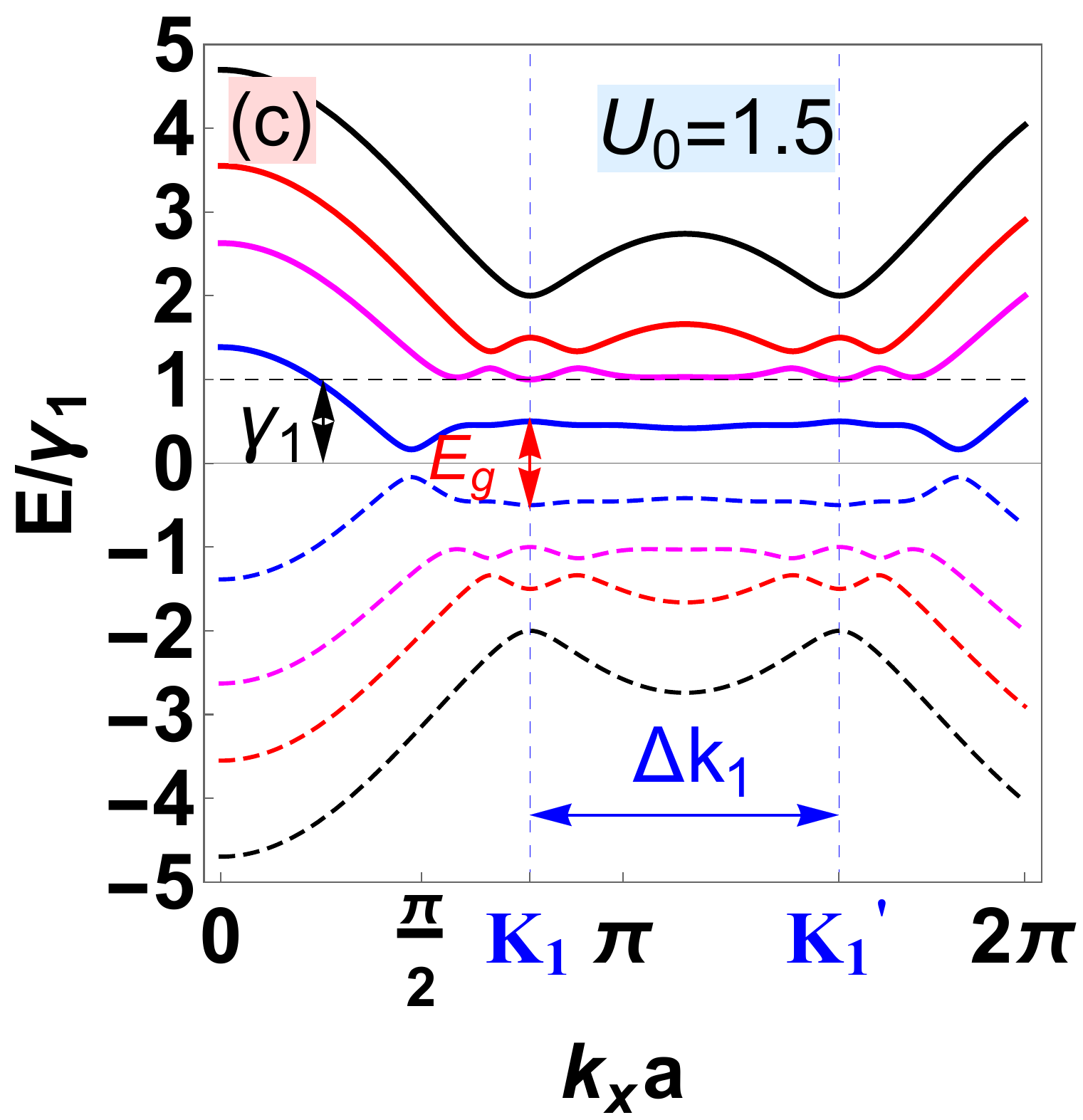}
\caption{(Color online) Band structure of ABCA-TTLG for $k_y=0$ and $t'=t$. The blue, magenta, 
red and black curves correspond to the tight-binding model. The solid curves correspond to 
the conduction bands and the dashed curves correspond to the valance bands.   (a) $U_{0}=0.3$, (b) 
$U_{0}=1$, (c) $U_{0}=1.5$. The vertical dashed line indicates the position of the Dirac point $K$.}\label{energyuk}
\end{figure}

In order to calculate the full dispersion of ABCA-TTLG we numerically diagonalize the Hamiltonian  
Eq. \eqref{HamABCA}. 
The resulting diagonal matrix 
then contains eight distinct entries corresponding to the energy bands compared with monolayer, 
AB-BL and ABC-TLG which contains two, four and six bands, respectively. 
These eight bands are plotted in Fig. \ref{energyeffec1}, \ref{EnergyGMKG} and \ref{energyuk}. 
In Fig. \ref{energyeffec1}, we show the typical behavior of the band structure dependence 
of next-nearest-neighbor  hopping $t'$ for fixed nearest neighbor $t$ and applied potential 
$U_{0}=0$. For $t'=t$ the energy vanishes at two points located at the Dirac corners 
$K_1$ and $K_{1}'$ of the hexagonal BZ spaced by $\Delta K_1=K_{1}'-K_1$ (see Fig. \ref{energyeffec1}(a)).
For $2t>t'>t$, the Dirac points approach each other, their distance varies as $\Delta K_2=K_{2}'-K_2<\Delta K_1$,  
the Dirac cones become anisotropic (deformed honeycomb lattice) (see Fig. \ref{energyeffec1}(b)). For $t'=2t$, 
the two Dirac points merge into a single point at $K_3=K_3'$ (see Fig. \ref{energyeffec1}(c)). For $t'>2t$, 
a gap opened $E_g\simeq0.694$eV between the conduction and valance bands, and the two Dirac points 
coincide at $K_4=K_4'$ (see Fig. \ref{energyeffec1}(d)).
 We notice that for $t'<2t$ and $U_{0}=0$  the band structure of ABCA-TTLG  in our Tight-binding model 
 consists of a set of four pairs cubic of bands, one of them touching each other at the points $K$ 
 and $K'$,  and the other three crossing at the energy $E=\pm \gamma_{1}$. These two bands
 intersect exactly at $E_F$ which corresponds to the special point $K$ in the reciprocal space 
 of the hexagonal lattice. The insets of Fig. \ref{energyeffec1} show the zoom of the bands at 
 the Dirac point and  the deformation of hexagonal Brillouin zone (lattice) by variation of
  the next nearest-neighbor hopping $t'$. 

For a more understanding of the effects of the next nearest neighbors $t'$ on band structure of 
ABCA-TTLG we show in Fig. \ref{EnergyGMKG} the $\pi$ bands for different values of $t'$ 
along the  lines $\Gamma-M-K-\Gamma$ in the first BZ. From this Figure it is very clear 
that the presence of the parameters $t'$ (as $t'$ increases) introduces a band gap on the band structure.

To illustrate the effects of the potential  $U_{0}$ 
on the band structure of ABCA-TTLG in the tight-binding model we plot it versus the transverse wave vector $k_x$
in Fig. \ref{energyuk} 
for some values 
the potential height 
$U_0=0.3, 1, 1.5$ with  
$t'=t$. The Bravais 
lattice retains its shape even if $U_{0}$ is strong but $U_{0}$ contribute to the opening of a 
gap between the conduction and valence bands. Note that  $\Delta K_1=K_{1}'-K_1$ remains 
unchangeable. More interesting features appear when zooming into the low energy regime 
around the point $K$  see Sec. \ref{Lowenergy}.

\begin{figure}[tb]
\centering
\includegraphics[width=5.8cm]{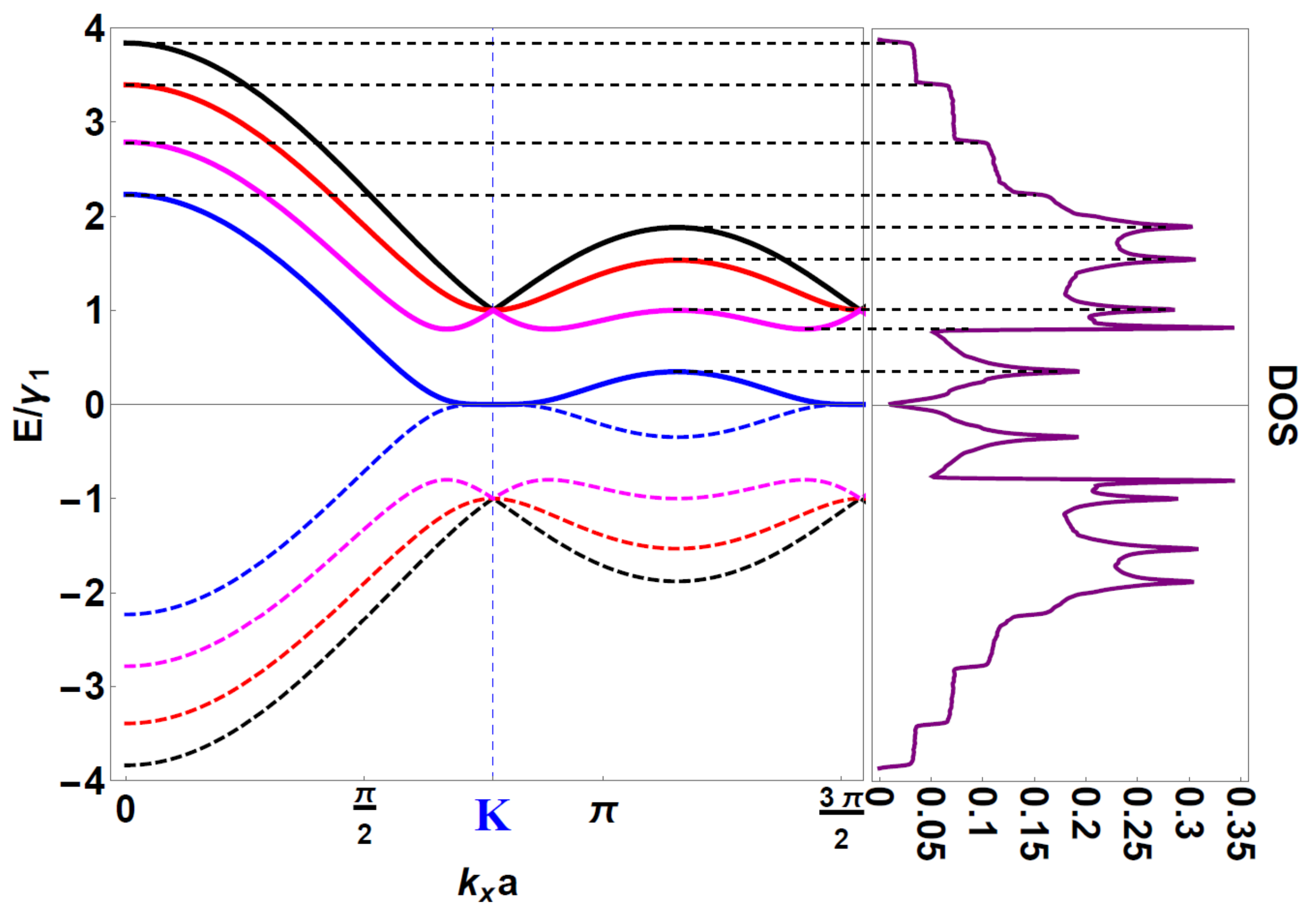}
\caption{(Color online) Band structure of ABCA-TTLG presented in 
Fig. \ref{energyeffec1}(a) from tight-binding model, with the right hand side panel 
exhibiting the corresponding DOS. Sharp spikes in the DOS are VHS. The positive VHS 
are indicated by the dashed lines. The energy dispersion relations are presented 
in the  energy interval $[-4\gamma_1, 4\gamma_1]$ in dimensionless units. The Fermi level is located at zero energy.}  
\label{dos}
\end{figure}

In Fig. \ref{dos}, the influence of next-nearest-neighbor $t'$ hopping on electronic DOS in 
ABCA-TTLG is shown. The  DOS have a zero value at the Fermi energy and ten sharp van Hove 
singularities (VHS) appear 
at the onset of each subband. We also see, as in the spectrum, an apparent symmetry for positive 
and negative  VHS in the DOS. This also implies zero-gap semiconductor with semi metallic 
behavior. In addition, sharp peaks in the DOS are observed, which are the characteristic 
signatures of the one-dimensional (1D) nature of conduction within a 1D system. We conclude 
that by varying $t'$, VHS can be brought to accessible energies, which is one of the 
important  features of the ABCA-TTLG.%
\subsection{Low-energy model}
\label{Lowenergy}
\begin{figure}[tb]
\centering
\includegraphics[width=2.9cm]{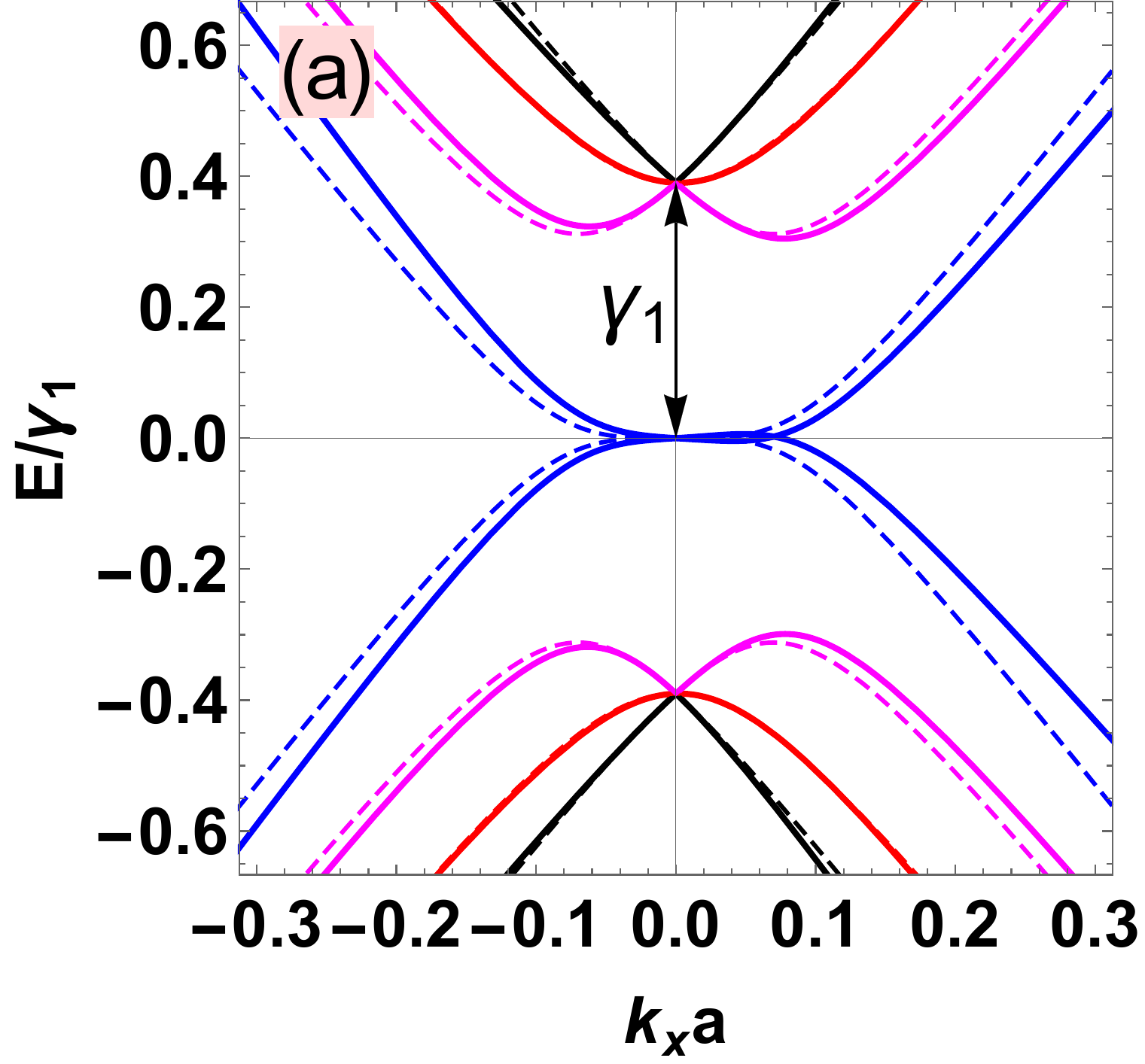}\includegraphics[width=2.9cm]{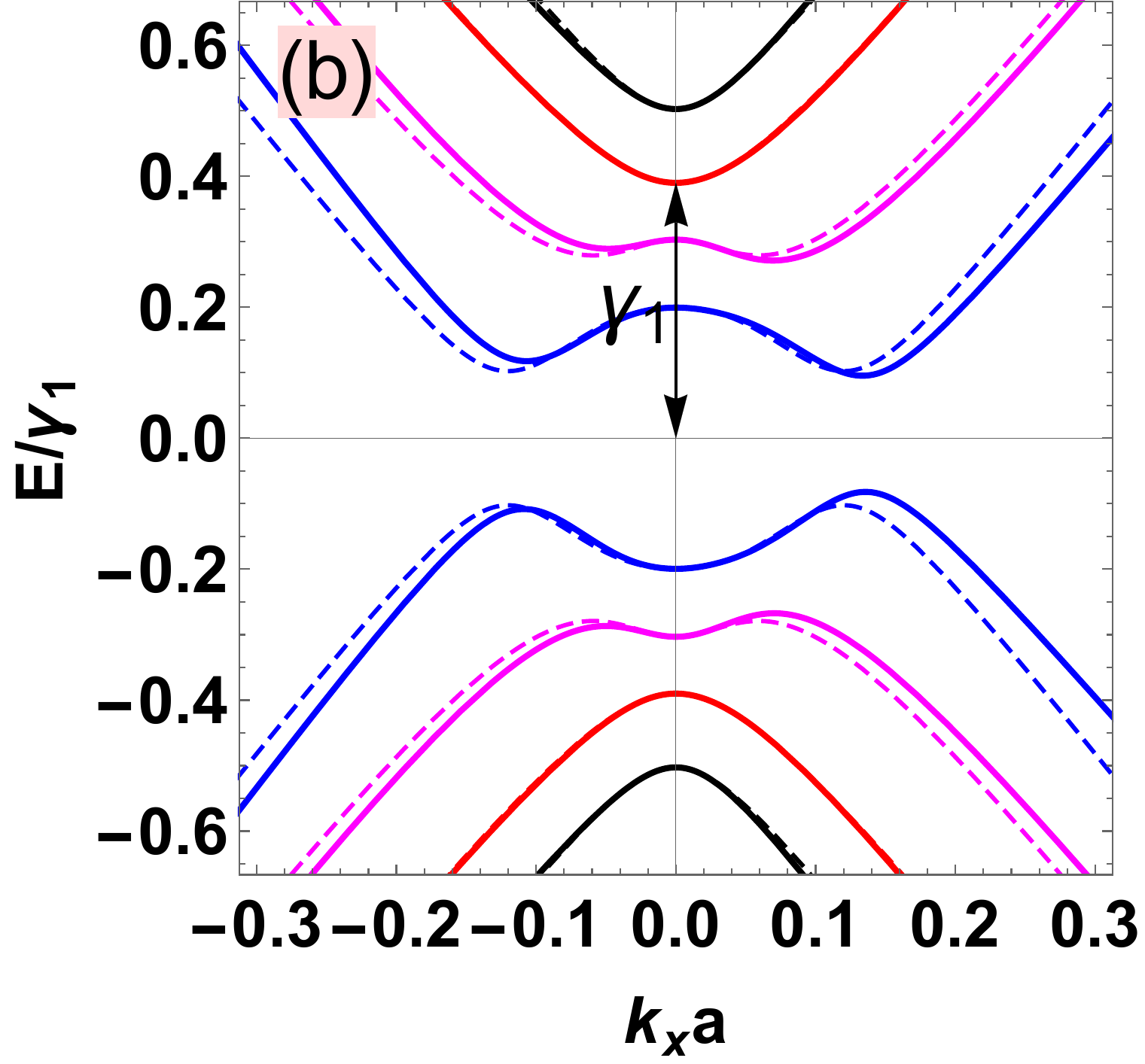}\includegraphics[width=2.81cm]{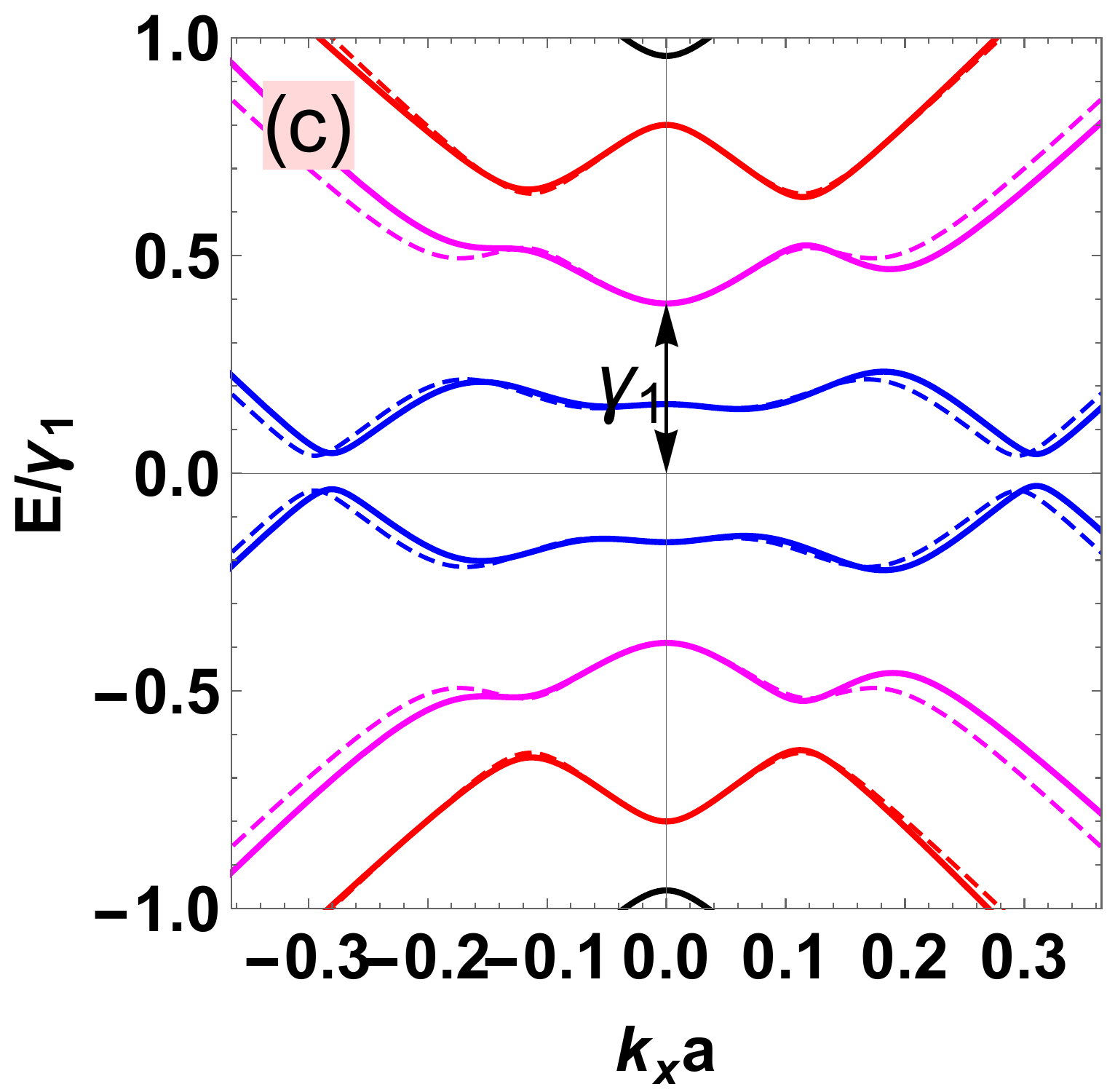}\\
\hspace{0.7cm}\includegraphics[width=1.3cm]{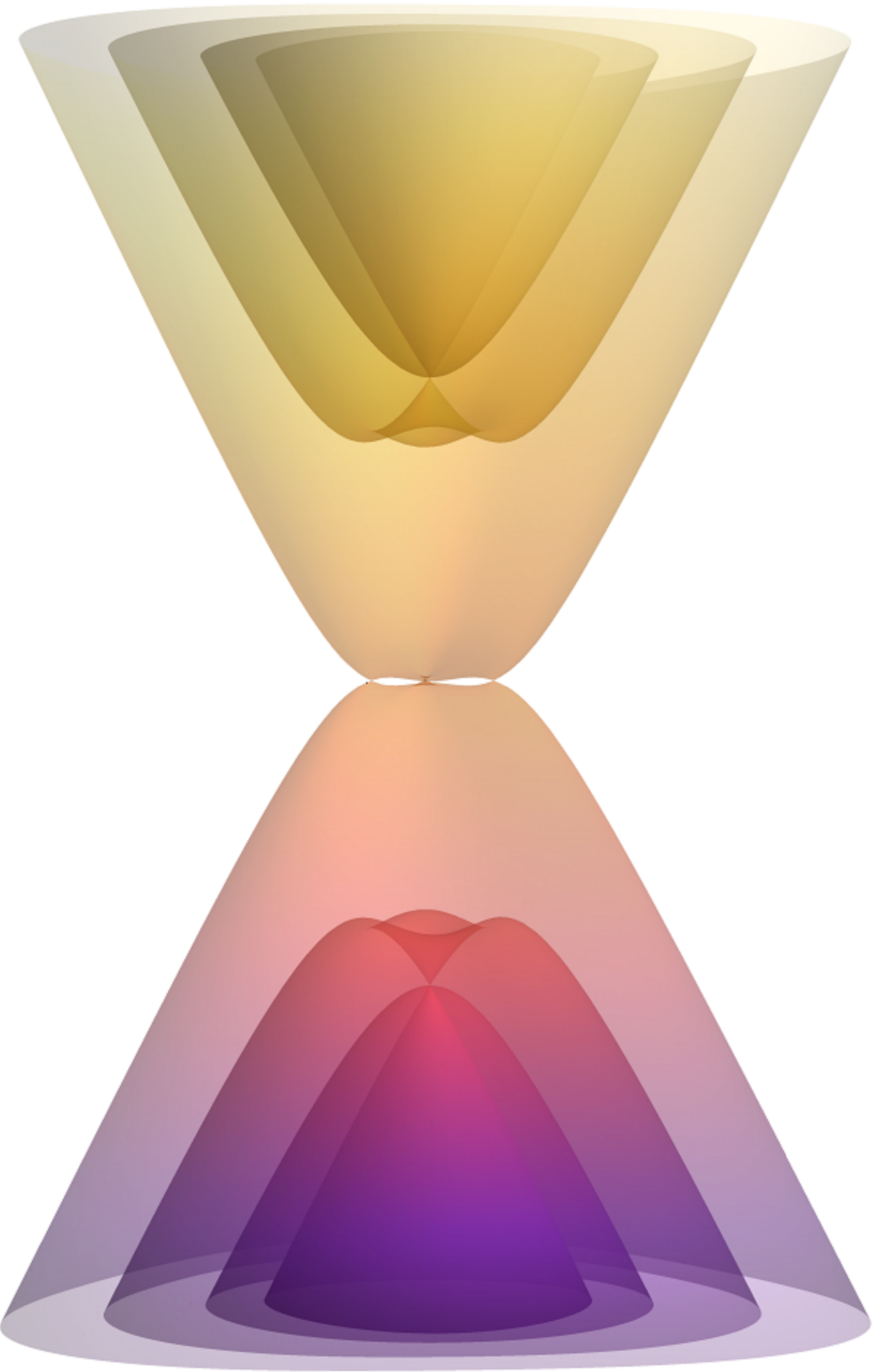}
\hspace{1.2cm}\includegraphics[width=1.4cm]{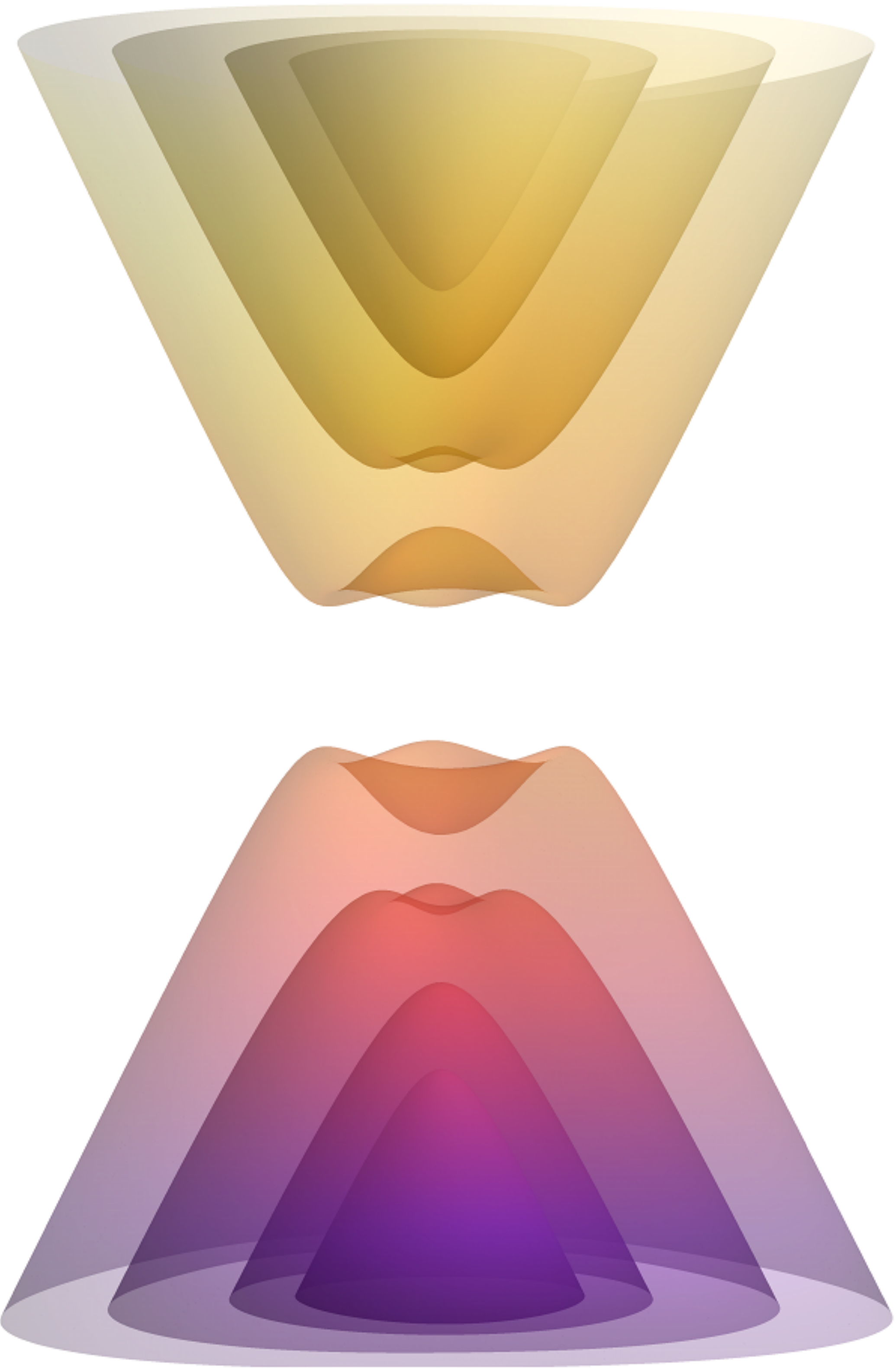}
\hspace{1.3cm}\includegraphics[width=1.45cm]{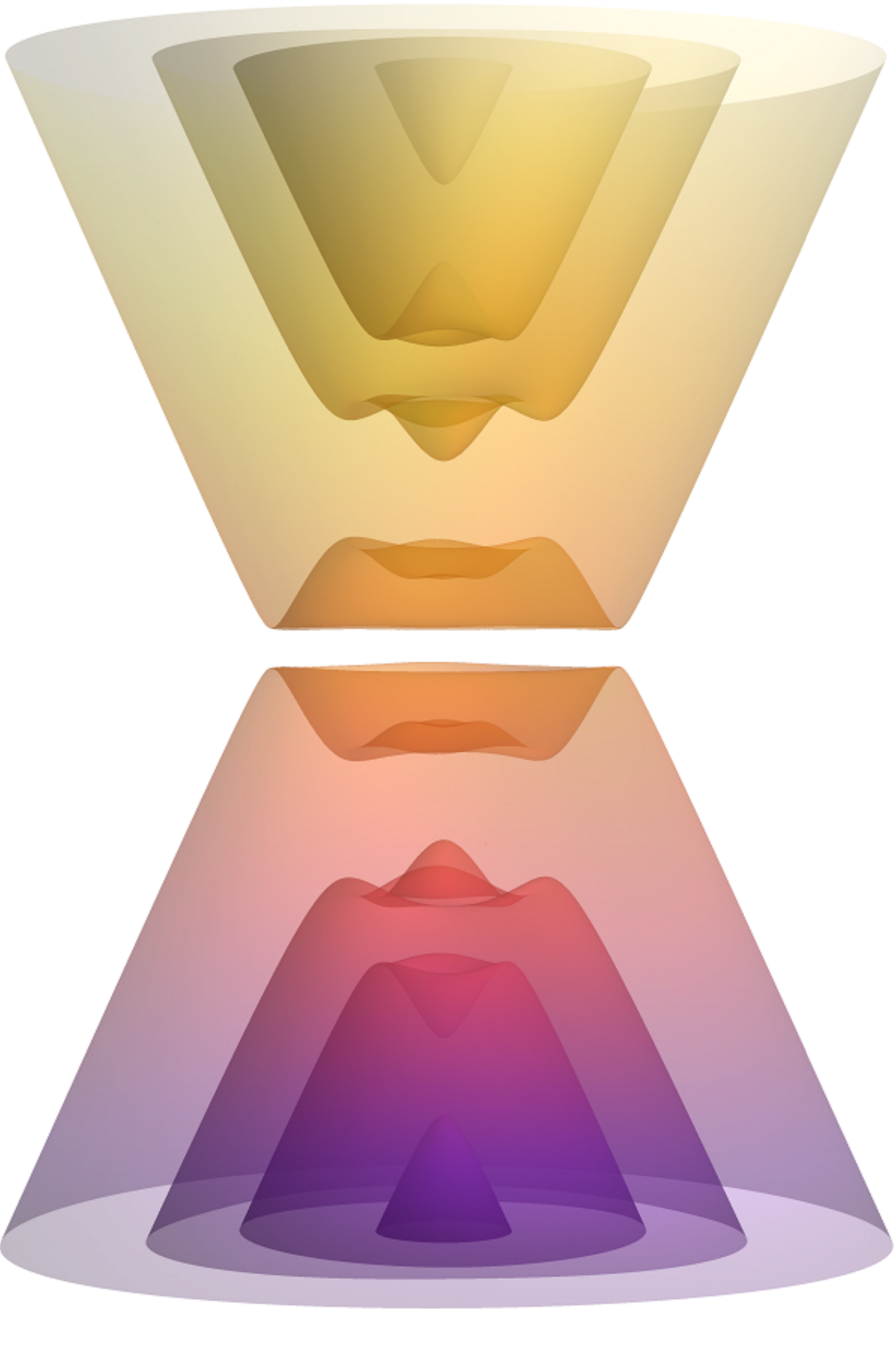}
\caption{(Color online) Band structure near the Dirac  point $K$, calculated in the 
low-energy model of ABCA-TTLG (top),  with the corresponding 3D (below), 
when the potential difference (a) $U_{0}=0$, (b) $U_{0}=0.2$ and (c) $U_{0}=0.8$. 
The solid curves correspond to the band structure in the first neighbor approximation 
(i.e., $\gamma_{0}$ and $\gamma_{1}$ only) and the dashed curves  correspond to the band 
structure in the full parameters model. A band gap is previously created by application 
of an external potential difference between top and bottom layers.}\label{EABCABkx}
\end{figure}
\begin{figure}[tb]
\centering
\includegraphics[width=8cm]{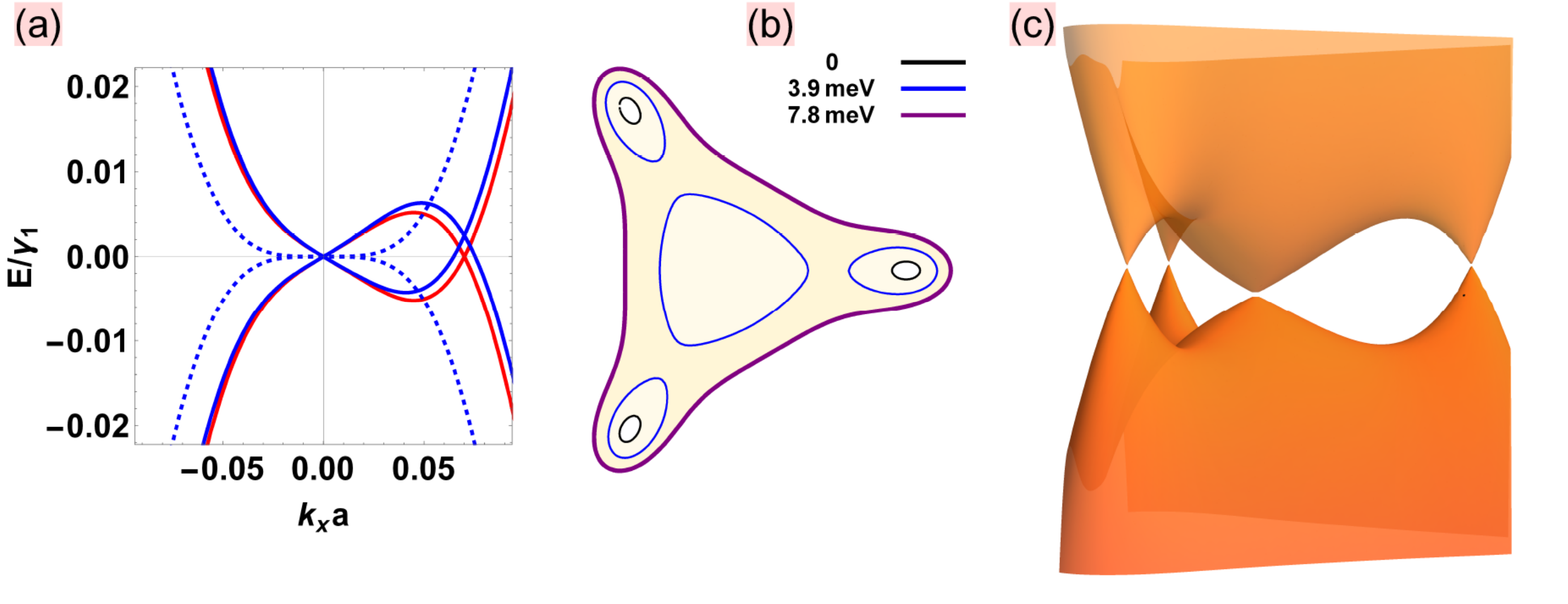}
\caption{(Color online) Energy spectrum of ABCA-TTLG near one of the Dirac  
point $K$ at low energy. The solid blue curve corresponds to the spectrum accounting for all 
interatomic hopping parameters considered in Fig \ref{stackingTTL}, the solid red curve corresponds 
to $\gamma_{4}=0$, and the dashed curve accounts only for $\gamma_{0}$ and $\gamma_{1}$.}\label{Trigonal}
\end{figure}

The band gap is an intrinsic property of semiconductors, which indeed hugely determines 
the transport and the optical properties. It should have a key role in modern device 
industry and technology. In this section, we focus on the sensitivity of the band structure
of ABCA-TTLG at low-energy by varying the 
applied 
potential $V$ defined in Eq. \ref{potential} 
between the top and the bottom layer. 
Using similar approach as before, we show in Fig. \ref{EABCABkx} the 2D (upper row) 
band structure at  point $K$ and the corresponding 3D plot (lower row) of the effective 
Hamiltonian derived from Eq. \ref{HamABCA} where $f(\mathbf{k})$ is given by Eq. \ref{hamDirac}. 
The potential is $\pm U_0$ for the top and bottom layer. The dashed curves also account 
for the skew hopping parameters defined in Fig. \ref{stackingTTL}(a) while the solid 
curve considers only nearest-neighbor interlayer hopping $t'$. It is clear that at high energy  
$E>6$meV the  effect of the skew hopping parameters $\gamma_3$ and $\gamma_4$ to 
these band structures depicted in Fig. \ref{EABCABkx} is negligible (see Fig. \ref{Trigonal}(a)). 
That is why we did not take into consideration their effects on the band gaps. 
Trigonal warping due to $\gamma_3$ becomes particularly relevant at very low energy 
of the order of $6$meV. Fig. \ref{Trigonal}(b)) shows contour plots of the lower electron 
band at $U_0=0$, showing that the band is trigonally warped, and the contour splits into 
three pockets at low energy. The detailed band structure and its relation to the band 
parameters will be studied in the present sections. 

\begin{figure}[tb]
\centering
\includegraphics[width=5.5cm]{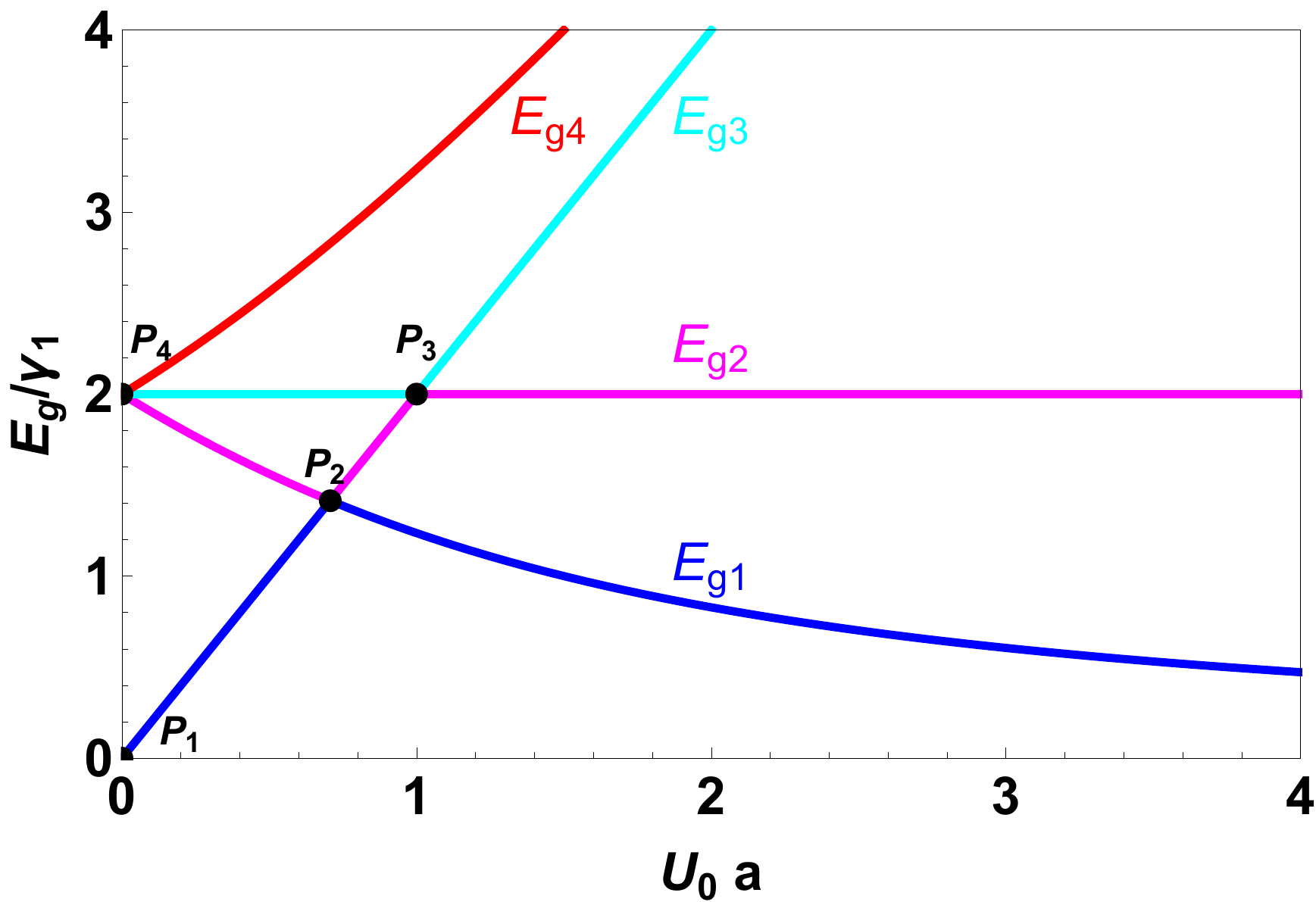}
\caption{(Color online) 
Band gap of ABCA-TTLG around the  Dirac
point $K$ for $k_y=0$ versus the potential difference $U_{0}$.}
\label{EGAP}
\end{figure}

For $U_{0}=0$  the spectrum consists of a set of four pairs cubic of bands, one of 
them touching each other at the  point $K$ and the other three crossing at an energy 
$E=\pm \gamma_{1}$ above (below) $K$, as shown in Fig. \ref{EABCABkx}(a). 
However, for $U_{0}\neq 0$ 
the location of the fundamental gap 
shifted, in  BZ, from $K$ to finite values of $K$, similar to what was observed 
in the gated multi-layer graphene 
\cite{Avetisyan200901,Avetisyan201032} 
(see Fig. \ref{EABCABkx}(b, c)). From these figures, four pairs of bands hybridize and 
repel each other and a sizable band gap is created at $K$, which could be 
opened and controlled by 
$\pm U_{0}$. 
The repelling of the two bands creates  band gaps $E_{gi}=E_{ci}-E_{vi}$ 
with $i=1,2,3,4$. We denote the pair of highest (lowest) bands near the Dirac point as 
the conduction (valence) bands. Around $K$  and for $k_y=0$, these bands exhibit
several local maxima and minima as the potential difference $U_{0}$ is increased 
(see Fig. \ref{EABCABkx}(b, c) and  Tab. \ref{table}). Always they are symmetric with 
respect to $E_F$. Therefore, the band gaps $E_{gi}$ are strongly dependent 
on $U_0$ 
and are given by 
\begin{eqnarray}
E_{g1}&=&\begin{cases}-U_{0}+\sqrt{U_{0}^{2}+4\gamma_{1}^{2}} & \mbox{for } U_{0}>\frac{\gamma_{1}}{\sqrt{2}} \\
2U_{0} & \mbox{for } U_{0}\leqslant\frac{\gamma_{1}}{\sqrt{2}}\label{egap1},
\end{cases}\\
E_{g2}&=&\begin{cases}2\gamma_1 & \mbox{for } U_{0}\geqslant\gamma_{1} \\
2U_{0} & \mbox{for } \frac{\gamma_{1}}{\sqrt{2}}<U_{0}<\gamma_{1}\\
-U_{0}+\sqrt{U_{0}^{2}+4\gamma_{1}^{2}} & \mbox{for } U_{0}\leqslant\frac{\gamma_{1}}{\sqrt{2}}\label{egap2},
\end{cases}\\
E_{g3}&=&\begin{cases}2U_{0} & \mbox{for } U_{0}\geqslant\gamma_{1} \\
2\gamma_1 & \mbox{for } U_{0}<\gamma_{1}\label{egap3},
\end{cases}\\
E_{g4}&=&\begin{cases}2U_{0} & \mbox{for all } U_{0}\label{egap4}
\end{cases},
\end{eqnarray}
These four bands are illustrated in Fig. \ref{EGAP} which clearly define a better understanding 
of the band structure of ABCA-TTLG dependence on the band gap. The gaps are measured in units of 
$\gamma_{1}$. Our main result shows that the first band gap $E_{g1}$ initially increases and falls 
as 
$U_0$ 
increased. $E_{g1}$  is zero at $U_{0}=0$ ($P_{1}$ point) and 
reaches a maximum $1.414$eV as $U_0$ reaches $0.707$V/\AA\ ($P_{2}$ point). 
The second band gap $E_{g2}$ exhibits one local minima at $U_{0}=0.707$V/\AA\ ($P_{2}$ point) 
and  two local maxima who have the same gap energy $2$eV at $U_0=0$ ($P_4$) and at $U_0=1$V/\AA\ ($P_3$), 
and then remains constant. $E_{g3}$ varies steadily until $U_{0}$ reaches $1$V/\AA\ ($P_3$) and then   
increases monotonically when $U_{0}$ increases. As long as $U_{0}$ increase the band gap 
$E_{g4}$ 
continuously increased. 

In general, for the ABCA-TTLG the coordinates of the points $P_{1}$, $P_{2}$, $P_{3}$ and $P_{4}$ 
illustrated in Fig. \ref{EGAP} are given by 
$P_{1}(0,0$), $P_{2}(\gamma_{1}/\sqrt{2},\sqrt{2}\gamma_{1})$,  $P_{3}(\gamma_{1},2\gamma_{1})$ 
and $P_{4}(0,2\gamma_{1}),$ which 
are strongly 
dependent  on gamma $\gamma_{1}$. 
We note that the behavior of the minimum band gap $E_{g1}$ in ABCA-TTLG is similar to 
ABC-TLG\cite{Kumar201101,Koshino2010}, which increases by increasing $U_0$ with band gaps 
below a critical value, and decreases with band gaps above this critical value. By contrast, 
in AB-BLG\cite{Koshino2010,Koshino200910} the situation is completely different because
$E_{g1}$ increases as $U_0$ increases.
\begin{table}[ht]
\begin{tabular}{|l|l|l|l|c|l|c|}
    \hline
    $\mathbf{U_{0}}$&$\mathbf{0}$ & $\mathbf{0.5}$ & $\mathbf{0.7}$ & $\mathbf{1}$ & $\mathbf{1.5}$ & $\mathbf{3}$\\
\hline
    $\mathbf{E_{g1}(ABCA)}$&$0$ & $1$ & $1.4$ & $1.236$ & $1$ & $0.605$\\\hline
\end{tabular}
\caption{Several values of the energy gap $E_{g1}$ for  a several values of $U_{0}.$}
\label{table}
\end{table}
\subsection{Two-band approximation model}
\label{Twoband}
\begin{figure}[tb]
\centering
\includegraphics[width=5.5cm]{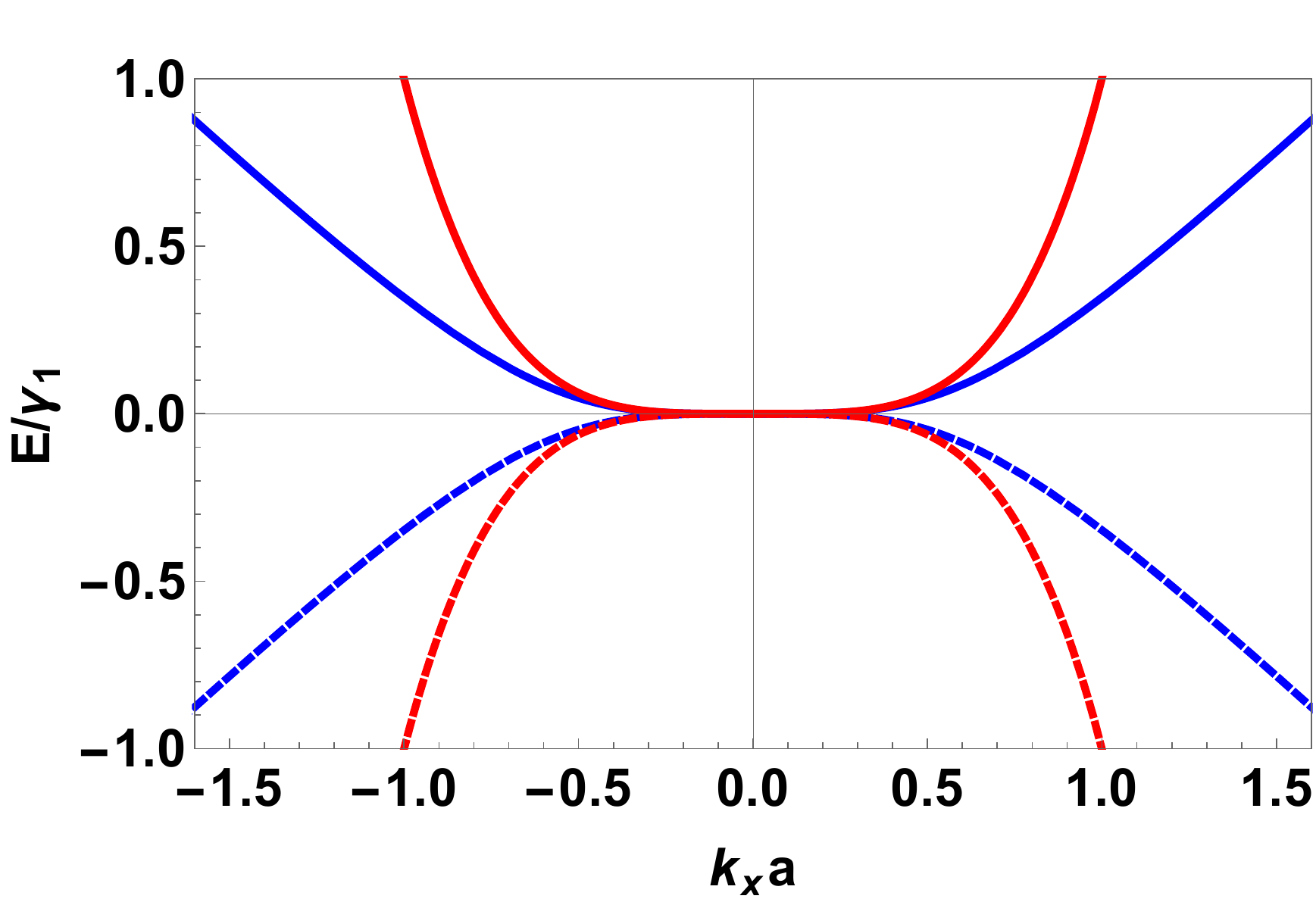}
\caption{Band structure of ABCA-TTLG around the Dirac point $K$ for $k_y=0$. 
The blue curves correspond to the low energy model (Eq. \ref{HamABCA}) and 
the red curves correspond to the two band approximation model (Eq. \ref{ReducedHamEq}).}
\label{energyeffec}
\end{figure}
The same as for AB-BLG and ABC-TLG, it is possible to introduce the effective two-band description 
at low energy approximation for the ABCA-TTLG. This two component Hamiltonian was derived in 
\cite{vanduppen2013, MacDonald2008, Nakamura} and is only valid as long as the electronic density 
is low enough, that is, when the Fermi energy is much smaller than $\gamma _{1}$ ($E\ll\gamma_{1}$).
We use similar analysis to that 
previously 
described in Ref.\cite{vanduppen2013}, which 
 yields the Hamiltonian\cite{Ben201301} 
\begin{eqnarray}
\mathcal{H}^{\prime}(\mathbf{k}) &=&\left[
\begin{array}{cc}
{V} & \alpha\left( k_{x}-ik_{y}\right) ^{4} \\
\alpha\left( k_{x}+ik_{y}\right) ^{4} & -V%
\end{array}%
\right],\label{ReducedHamEq}
\end{eqnarray}
where $\alpha=\frac{\left(\hbar v_{F}\right)^{4}}{\gamma
_{1} ^{3}}$. In the two band approximation model, the Hamiltonian in Eq. \ref{ReducedHamEq} 
has a plane wave solution given by two propagating waves, one right and one left moving, and 
six evanescent waves. The wave vectors of these plane waves are the solutions of the equation
\begin{equation}
\alpha^{2}\left( k_{j}^{2}+k_{y}^{2}\right)^{4}+V^{2}=\varepsilon ^{2},
\end{equation}%
where $k_{y}=\varepsilon^{1/4}\sin(\phi _{k})$ is the transverse wave vector, 
$\phi _{k}=\arctan \left( k_{y}/k_{x}\right) $ is the angle of the wave vector 
${\mathbf{k}}$ with the normal chosen perpendicular to the pn junction. 
The solution of this equation is $\pm k_{j}$ with four different values for $k_{j}$ and the dispersion relation is given by 
\begin{equation}
\varepsilon _{\pm}=\pm\sqrt{\alpha^{2}\left( k_{j}^{2}+k_{y}^{2}\right)^{4}+V^{2}}.
\end{equation}%
The validity of this approximation is presented in Fig. \ref{energyeffec} 
where we 
show the dispersion relation obtained in the low energy model (blue curves ) and the dispersion 
relation obtained from the Hamiltonian in the two bands model (red curves). We clearly see that 
for $E\ll \gamma_1$ the two band spectrum for the low-energy model are in good agreement with 
the two band spectrum for low energy Eq. \ref{ReducedHamEq}.
\section{Transmission probability}
\label{trans}
In this paragraph, we focus on the transmission probability through a pn junction 
in the two band approximation model (Eq. \ref{ReducedHamEq}) and   adopt the same 
approach used by Ben {\it et al.} \cite{Ben201301}. The eigenstates of these $2\times2$ Hamiltonians 
are eight-component spinors consisting of a superposition of four times two oppositely 
propagating or evanescent waves characterized by four distinct wave vectors 
denoted as $k_{1}$, $k_{2}$, $k_{3}$ and $k_{4}$. For a system that is translational 
invariant in the $y$ direction, the energy and $k_{y}$ dependence on these wave vectors $k_{i}$ can be found from
\begin{eqnarray}
\text{det}\left[\mathcal{H}^{\prime}(\mathbf{k})-EI_{8}\right]=0\label{eigens},
\end{eqnarray}
and $\mathcal{H}(\mathbf{k})$ is given by Eq. \ref{ReducedHamEq}.
The eigenstates solution 
can be written as a product of matrices,
\begin{equation}
\psi\left( x,y\right)=\mathcal{MW}\left( x,y\right)\mathcal{C},\label{EIGENSTATES}
\end{equation}
where $\mathcal{M}$ is a $8\times2$ matrix expressing the relative importance of 
the different components of the spinor that can be constructed by solving the Dirac equation  
$\mathcal{H}(\mathbf{k})\psi\left( x,y\right)=E\psi\left( x,y\right)$. The matrix $\mathcal{M}$ is given by
\begin{equation}
\mathcal{M} =  \left[
\begin{array}{cccccccc}
1 & 1 & 1 & 1 & 1  & 1 & 1 & 1 \\
f_{1}^{+} & f_{1}^{-} & f_{2}^{+}& f_{2}^{-} & f_{3}^{+} & f_{3}^{-} & f_{4}^{+} & f_{4}^{-}%
\end{array}%
\right],
\end{equation}
where $f_{j}^{\pm}=\frac{\left(
\pm k_{j}+ik_{y}\right) ^{4}}{\varepsilon}$
and the matrix $\mathcal{W}$ 
is 
\begin{widetext}
\begin{equation}
\mathcal{W}\left( x,y\right)=\text{diag}\left[
e^{ik_{1}x}\ e^{-ik_{1}x},e^{ik_{2}x},e^{-ik_{2}x},e^{ik_{3}x},e^{-ik_{3}x},e^{ik_{4}x},e^{-ik_{4}x}\right] e^{ik_{y}y},
\end{equation}
\end{widetext}
due to the translational symmetry in the $y$ direction, the $y$ 
dependency is incorporated in an exponential phase factor and will be ignored 
from this point on. We denote the four component vector $\mathcal{C}$ as
\begin{equation}
\mathcal{C} =  \left[ a_{1}^{+},a_{1}^{-}, a_{2}^{+},a_{2}^{-}, a_{3}^{+},a_{3}^{-}, a_{4}^{+},a_{4}^{-}\right]
^{T},
\end{equation}
where the subscript $i=1, \cdots 4$ refers to the corresponding wave vector and the superscript plus/minus indicates 
the right/left propagating or evanescent states. The boundary conditions of the system under consideration 
will determine which of the components of vector $\mathcal{C}$ are zero. To find the transmission 
probability for a pn junction, one has to equate the plane wave solutions and all the derivatives 
up to $3^{th}$ order of the region before the junction (region I) with those of the region behind 
it (region II) at the junction's edge at $x=0$, giving rise to 
a set of four two component equations
\cite{Ben201301}
\begin{equation}
\left\{
\begin{array}{llll}
\mathcal{M}_{I}\mathcal{W}_{I} \mathcal{C}_{I}=\mathcal{M}%
_{II}\mathcal{W}_{II} \mathcal{C}_{II} \\
\mathcal{M}_{I}\frac{\partial \mathcal{W}_{I} }{\partial x}%
\mathcal{C}_{I}=\mathcal{M}_{II}\frac{\partial \mathcal{W}_{II}}{\partial x}\mathcal{C}_{II} \\
\mathcal{M}_{I}\frac{\partial^{2}\mathcal{W}_{I} }{%
\partial x^{2}}\mathcal{C}_{I}=\mathcal{M}_{II}\frac{\partial^{2}%
\mathcal{W}_{II}}{\partial x^{2}}\mathcal{C}_{II}  \\
\mathcal{M}_{I}\frac{\partial^{3}\mathcal{W}_{I} }{%
\partial x^{3}}\mathcal{C}_{I}=\mathcal{M}_{II}\frac{\partial^{3}%
\mathcal{W}_{II}}{\partial x^{3}}\mathcal{C}_{II},
\end{array}%
\right.
\end{equation}%
where the matrix $\mathcal{W}$ is evaluated at $x=0$. This leads to the matrix
\begin{widetext}
\begin{equation}
\mathcal{M} =  \left[
\begin{array}{cccccccccc}
1 & 1 & 1 & 1 & 1  & 1 & 1 & 1\\
f_{1}^{+} & f_{1}^{-} & f_{2}^{+}& f_{2}^{-} & f_{3}^{+} & f_{3}^{-} & f_{4}^{+} & f_{4}^{-} & \\
ik_{1} & -ik_{1} & ik_{2} & -ik_{2} & ik_{3}  & -ik_{3} & ik_{4} & -ik_{4} & \\
ik_{1}f_{1}^{+} & -ik_{1}f_{1}^{-} & ik_{2}f_{2}^{+}& -ik_{2}f_{2}^{-} & ik_{3}f_{3}^{+} & 
-ik_{3}f_{3}^{-} & ik_{4}f_{4}^{+} & -ik_{4}f_{4}^{-} & \\
-k_{1}^{2} & -k_{1}^{2} & -k_{2}^{2} & -k_{2}^{2} & -k_{3}^{2}  & -k_{3}^{2} & -k_{4}^{2} & -k_{4}^{2} \\
-k_{1}^{2}f_{1}^{+} & -k_{1}^{2}f_{1}^{-} & -k_{2}^{2}f_{2}^{+}& -k_{2}^{2}f_{2}^{-} & -k_{3}^{2}f_{3}^{+} 
& -k_{3}^{2}f_{3}^{-} & -k_{4}^{2}f_{4}^{+} & -k_{4}^{2}f_{4}^{-} \\
-ik_{1}^{3} & ik_{1}^{3} & -ik_{2}^{3} & ik_{2}^{3} & -ik_{3}^{3}  & ik_{3}^{3} & -ik_{4}^{3} & ik_{4}^{3}\\
-ik_{1}^{3}f_{1}^{+} & ik_{1}^{3}f_{1}^{-} & -ik_{2}^{3}f_{2}^{+}& ik_{2}^{3}f_{2}^{-} & -ik_{3}^{3}f_{3}^{+} 
& ik_{3}^{3}f_{3}^{-} & -ik_{4}^{3}f_{4}^{+} & ik_{4}^{3}f_{4}^{-}
\end{array}%
\right].
\end{equation}
\end{widetext}
Normalizing the incident wave on the right propagating wave before 
the junction by putting $a_{1,I}^{+}=1$ and applying boundary conditions 
$a_{j,I}^{-}=a_{j,II}^{+}=0$ for $j\neq 1$ to suppress the non 
normalizable plane wave functions. Therefore, the transmission $T$ 
and the reflection probabilities $R$ are given by
\begin{equation}
T=\left\vert a_{j,II}^{+}\right\vert ^{2}, \qquad 
R=\left\vert
a_{j,I}^{-}\right\vert ^{2}.
\end{equation}%
\begin{figure}[tb]
\centering
\includegraphics[width=2.5cm]{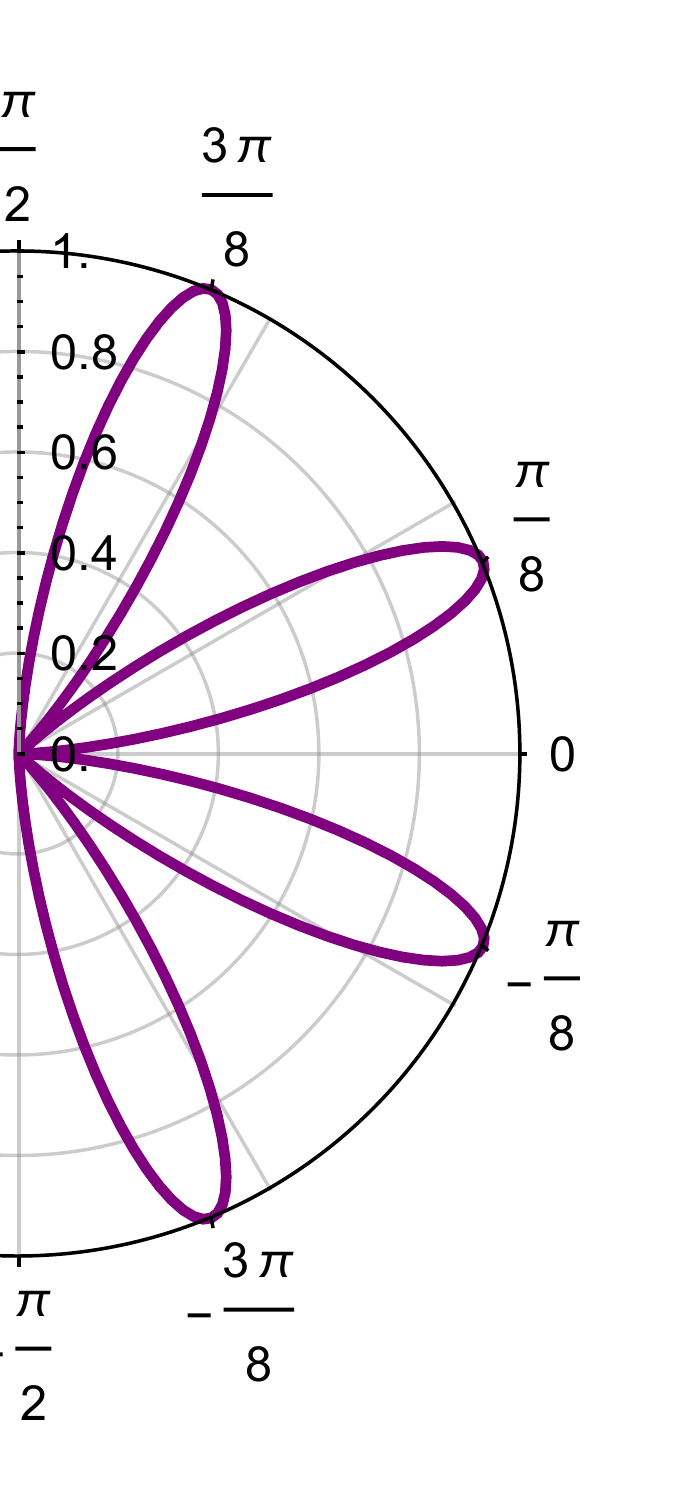}\\
\includegraphics[width=3.81cm]{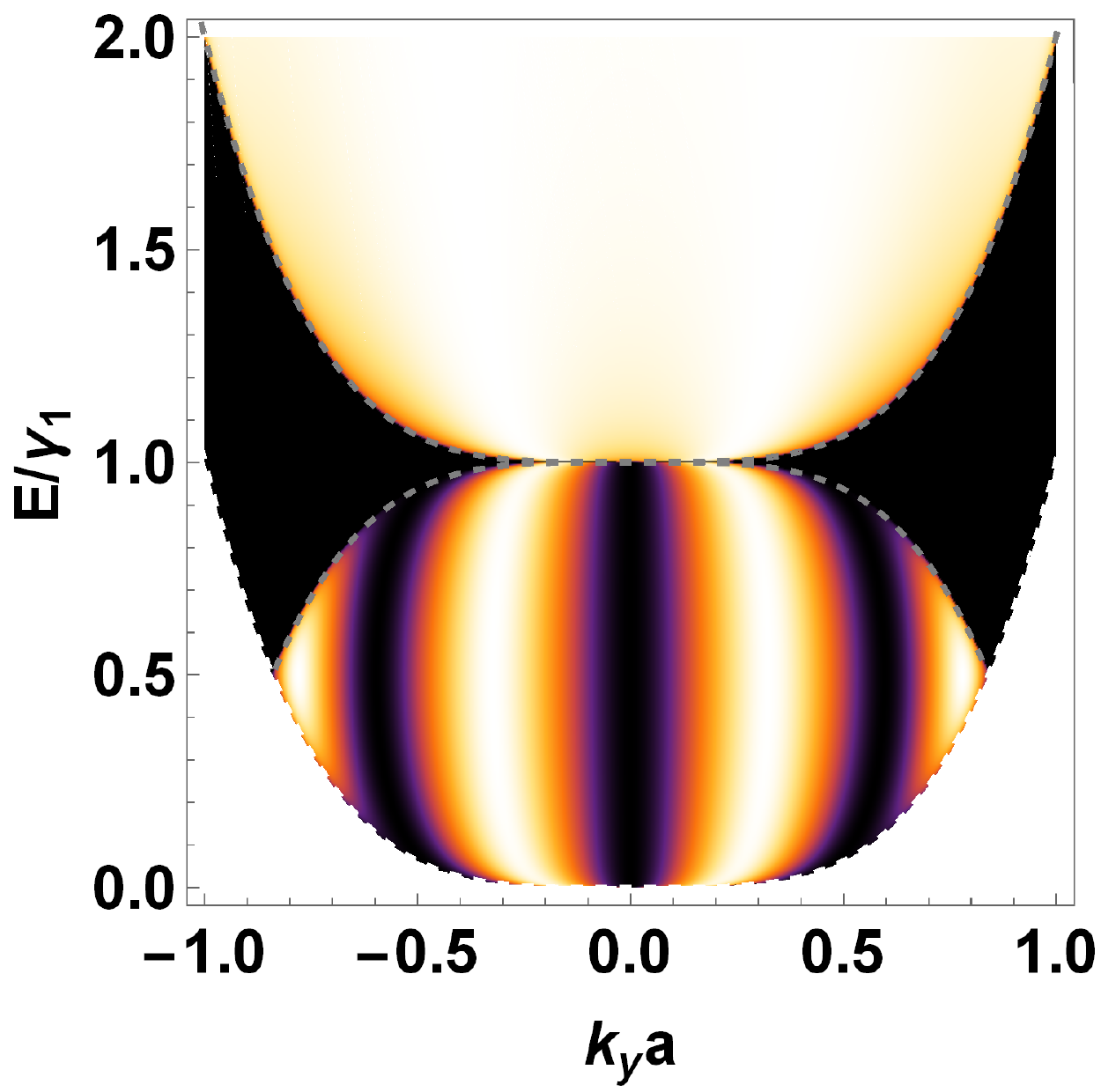}
\includegraphics[width=4.7cm]{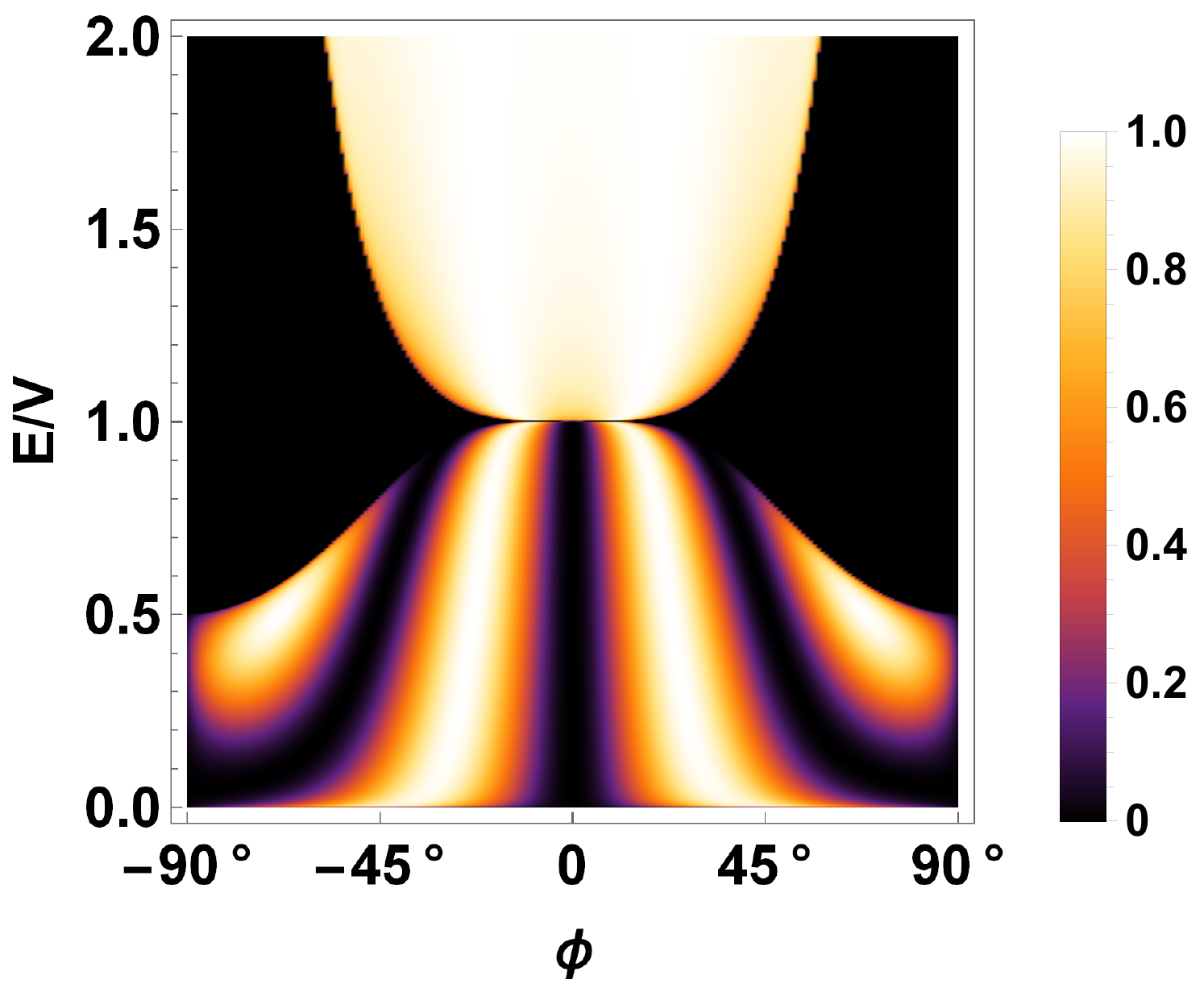}
\caption{(Color online) Transmission probability through a pn junction of height $V=2E$ 
in ABCA--TTLG using the two band Hamiltonian in Eq. \ref{ReducedHamEq}. The top row shows 
the  transmission dependence on the angle. The bottom row shows the counterplots of the transmission 
versus of the energy, 
transverse wave vector and 
angle.}
\label{Polars}
\end{figure}
The numerical results of the transmission probability for ABCA-TTLG is depicted in 
Fig. \ref{Polars}. At normal incidence, i.e. $ky = 0$ ($\phi=0$), 
the transmission equals zero independently of energy or height of the step. 
There are four high-transmission regions for $E<V$ (see Fig. \ref{Polars} bottom row), 
which are typical for tetralayer graphene, shifted away from each other and each one forms a Klein (KT) 
and anti-Klein (AKT) tunneling region. The expected angles  ($\phi=\pm \pi/8, \pm 3\pi/8$) 
for KT and  ($\phi=0, \pm\pi/4$) for ATK (see Fig. \ref{Polars} top row), which are in agreement
with those obtained by Ben {\it et al.} \cite{Ben201301}
%
\section{Conclusion}
\label{Conclusion}
In conclusion, we have analyzed the band structure, DOS and transmission in ABCA-TTLG. Tight 
binding Hamiltonian containing nearest-neighbors $t$ and $t'$, effective Hamiltonian and two 
band approximation Hamiltonian with interlayer potential difference parameters have been employed. 
The expressions of band gaps  around the $\mathbf{k}$ are obtained. Interlayer potential difference 
$U_0$ is found to be responsible for generating  band gaps near Dirac point in ABCA-TTLG. The next 
nearest-neighbor hopping $t'$ breaks the symmetry of Bravais lattice and the corresponding BZ 
(e.g. its high symmetry points such as corners $K$ and $K'$ move) and the Dirac points may 
merge and move away from the high symmetry points. This gives rise to 1D-like VHS in the  DOS. 
Using the two-band approximation model, the KT and AKT angles are  obtained.
\section{Acknowledgement}
The generous support provided by the Saudi Center for Theoretical
Physics (SCTP) is highly appreciated by all authors.

\end{document}